\documentclass[]{bytedance_seed}

\usepackage[toc,page,header]{appendix}

\usepackage{minitoc}

\usepackage{makecell}
\usepackage{verbatim}
\usepackage{multirow}
\usepackage{tcolorbox}
\usepackage{algorithm}
\usepackage{algorithmic}
\usepackage[table]{xcolor}

\title{Designing quantum chemistry algorithms with just-in-time compilation}

\author[1, \dagger]{Xiaojie Wu}
\author[1]{Qiming Sun}
\author[1]{Yuanheng Wang}

\affiliation[1]{ByteDance Seed}

\contribution[\dagger]{Corresponding authors}

\abstract{
We introduce just-in-time (JIT) compilation to the integral kernels for Gaussian-type orbitals (GTOs) to enhance the efficiency of electron repulsion integral computations. For Coulomb and exchange (JK) matrices, JIT-based algorithms yield a 2× speedup for the small 6-31G* basis set over GPU4PySCF v1.4 on an NVIDIA A100-80G GPU. By incorporating a novel algorithm designed for orbitals with high angular momentum, the efficiency of JK evaluations with the large def2-TZVPP basis set is improved by up to 4×. The core CUDA implementation is compact, comprising only \textasciitilde1,000 lines of code, including support for single-precision arithmetic. Furthermore, the single-precision implementation achieves a 3× speedup over the previous state-of-the-art.
}

\date{\today}
\correspondence{Xiaojie Wu at \email{xiaojie.wu@bytedance.com}}

\checkdata[Project Page]{\url{https://github.com/ByteDance-Seed/JoltQC.git}}

\begin{document}
\maketitle


\section{Introduction}
Traditional quantum chemistry software \cite{GPU4PySCF1,GPU4PySCF2,terachem2024} has historically relied on ahead-of-time (AOT) compilation and monolithic code generation, with tightly coupled Fortran or C++ implementations \cite{ORCA5,qchem5,g16} tuned for a fixed set of use cases. Although this design has served well for decades, it poses serious limitations in some modern high-performance computing (HPC) environments where GPUs are the primary source of computational power. Algorithms in electronic structure theory often require handling a combinatorial variety of tensor contraction patterns and sparsity conditions that depend on the input molecular system. Static compilation must conservatively support all conditional branches, parameter combinations, and numerical thresholds, leading to bloated binaries, excessive control flow divergence, poor cache and register utilization, and inefficient use of memory bandwidth. On GPUs, these inefficiencies are further magnified by strict constraints on register pressure, warp divergence, and hardware occupancy, often requiring significant manual optimization. For example, most quantum chemistry packages deal with low-angular-momentum integrals successfully but still struggle with the performance of high-angular-momentum integrals, which suffer from all of these problems.

Just-in-time (JIT) compilation offers a transformative solution by enabling runtime code specialization based on the actual inputs: e.g., integral angular-momentum pattern, the number of primitives in an orbital shell, and task type. By generating code on demand, JIT techniques can eliminate irrelevant branches, tailor memory access patterns, and fuse computational stages that would be separate in statically compiled workflows. In GPU architectures, this dynamic specialization translates into reduced register pressure, improved memory coalescing, and better occupancy. Similar ideas have seen widespread adoption in the machine learning community, where frameworks such as PyTorch's TorchScript \cite{pytorch}, Triton \cite{triton} and JAX \cite{jax2018github} leverage JIT to fuse operations, reduce memory movement, and generate hardware-specific kernels. These successes demonstrate the maturity and impact of JIT compilation as a general-purpose performance strategy in high-throughput numerical computing, and motivate its application to quantum chemistry workloads.

Beyond runtime performance, just-in-time compilation also offers substantial advantages to developers. JIT facilitates rapid prototyping and testing: since code is generated at runtime, one can easily benchmark different kernel variants, tiling strategies, or instruction sequences, without recompiling and linking the entire application. This accelerates development cycles, simplifies debugging, and lowers the barrier to contributing performance improvements across architectures. Furthermore, 
developers can explore specialized algorithms that would be
impractical to hand-code under the AOT model. In this paper, we demonstrate several algorithms due to this benefit. The development experience can be even further improved: Within a narrow area of application such as machine learning, SQL and GPU programming, one can design domain-specific languages (DSLs) via LLVM \cite{llvm}. Frameworks like TensorFlow\cite{tensorflow2015-whitepaper}, PyTorch\cite{pytorch}, JAX\cite{jax2018github}, and TVM incorporate their own DSLs—such as TensorFlow's GraphDef and XLA HLO, PyTorch's TorchScript and FX, JAX's JAXpr, and TVM's TIR—to represent algorithms in intermediate forms that enable aggressive optimization and efficient execution. These DSLs allow developers to write high-level code while benefiting from low-level performance tuning through JIT compilation and backend-specific code generation, bridging the gap between productivity and performance.

The Schwarz inequality provides rigorous upper bounds on integral magnitudes, enabling most shell quartets to be evaluated in single precision (FP32) rather than the conventional double precision (FP64) \cite{terachem-mixed,quick-mixed}. JIT compilation makes this mixed-precision approach particularly easy to implement: since precision is a compile-time parameter, the same source code generates both FP32 and FP64 kernels without code duplication.
Most integral tasks can be calculated in single-precision as compared to the commonly-used double-precision. Single precision algorithms benefit quantum-chemistry computations in three key ways: (1) deep learning inference GPUs and consumer-grade GPUs typically feature far more FP32 units than FP64 units, consume less power, and are more readily available; (2) single-precision data occupy half the storage, allowing more values to reside in shared memory and registers, improving data locality and throughput; and (3) hardware-accelerated FP32 implementations of the exponential (exp) and error function (erf) offer 4×–8× speedups, alleviating what is a computational bottleneck in certain workloads. 
JIT compilation exploits all three advantages automatically: the same kernel source is compiled once for FP32 and once for FP64, with the compiler selecting the
  appropriate hardware intrinsics and register allocation for each precision — no manual code duplication required.
  
{\bf Contributions}: 1) We introduce JIT to the quantum chemistry community and optimize the two-electron integral kernels (1q1t in Sec. \ref{sec:1t1q} and 1qnt in Sec. \ref{sec:1qnt}) with compilation techniques. Our framework integrates seamlessly with the quantum chemistry package GPU4PySCF \cite{GPU4PySCF1, GPU4PySCF2} and is released as the open-source JoltQC library. 2) We develop a novel fragmentation algorithm for high angular momentum integrals, improving data locality and alleviating memory-bandwidth bottlenecks through multilevel reduction. 3) We benchmark single-precision integral kernels under our JIT framework and demonstrate substantial performance gains when combining JIT compilation with our new algorithmic strategies.
\section{A short formulation of Rys quadrature}
Evaluation of the electron repulsion integral is one major bottleneck of density functional theory (DFT), the most widely used electronic structure theory \cite{DFT_overview}. In this paper, the two-electron repulsion integrals $(ij|kl)$ are calculated with the Rys quadrature algorithm \cite{rys_quadrature}. Other popular integral algorithms such as McMurchie–Davidson \cite{MD_original} and Head-Gordon-Pople \cite{HGP_original} have been implemented in \cite{LibintX,terachem2024,Gamess_f}. Reimplementation of these methods with JIT is out of the scope of this paper. In this section, we briefly formulate the Rys quadrature algorithm for clarifying the notations and data dependency used in the rest of the paper. For detailed derivation and numerical stability analysis, we refer the readers to \cite{rys_quadrature}.

Here we are using $ijkl$ notations instead of conventional $\mu\nu\lambda\sigma$ notations to index atomic orbitals, to be consistent with the pseudo-code later in this paper. Since each of the Gaussian type orbitals is contracted from a set of primitive Gaussian functions \cite{purple_book_chapter_basis}, the integral tensor for a shell quartet is expressed as the sum of the primitive integral tensors
\begin{equation}
\label{eq:rys0}
(ij|kl)
=
\sum_{p\in i}\sum_{q\in j}\sum_{r\in k}\sum_{s\in l}
c_{ip}\,c_{jq}\,c_{kr}\,c_{ls}\;
(pq|rs).
\end{equation}
$c_{ip}$, $c_{jq}$, $c_{kr}$, and $c_{ls}$ are the pre-defined contraction coefficients for each shell in the basis set. The number of Gaussian functions in each primitive shell is $(l+1)(l+2)/2$ or $2l+1$ for cartesian atomic orbitals or spherical atomic orbitals respectively, where $l$ is the angular momentum of the shell. Only Cartesian orbitals are considered in the following sections, and spherical orbitals are just linear combinations of Cartesian orbitals with fixed transformation coefficients. These linear transformations can be performed efficiently outside the kernel. As a result, each dimension of the primitive integral $(pq|rs)$ is $N^f = (l+1)(l+2)/2$. We denote the dimension of the integral for each shell quartet as $(N_i^f, N_j^f, N_k^f, N_l^f)$.

Within Rys quadrature formalism, each primitive integral can be evaluated as a weighted sum of the integrant at specific quadrature root points
\begin{equation}
\label{eq:rys1}
    (pq|rs)=\sum_{r}^{N_r} (pq|rs)(t_r) \cdot w_r\\
\end{equation}
where $
N_{\text{r}}
=
\left\lfloor (l_i + l_j + l_k + l_l)/2 \right\rfloor + 1 $ is the number of Rys roots.
 $t_r$ and $w_r$ are the roots and weights of Rys quadrature. The roots and weights depend on the exponents and center (atomic) positions of the four primitive Gaussian functions, the total angular momentum, and a range-separation parameter $\omega$ if needed. The calculation of Rys roots and weights relies on interpolations from tabulated data, and the details have been documented in \cite{rys_quadrature}. 
 The number of Rys roots is determined at the compilation time, and only the data~\cite{GPU4PySCF2,libcint} associated with the number of roots are compiled into the binary. The primitive integral at each root position can be factored into intermediate variables in three spatial directions:
\begin{equation}
\label{eq:rys2}
\begin{aligned}
    (pq|rs)(t_r) = &I_x(a_x, b_x, c_x, d_x, t_r) \cdot I_y(a_y, b_y, c_y, d_y, t_r) \\
    & \cdot I_z(a_z, b_z, c_z, d_z, t_r)\\
     \text{for any } & (p, a_x,a_y,a_z) \in S_{l_i}, (q, b_x,b_y,b_z) \in S_{l_j}, \\
    & (r, c_x,c_y,c_z) \in S_{l_k}, (s, d_x,d_y,d_z) \in S_{l_l}\\
\end{aligned}
\end{equation}
$I_x$, $I_y$, and $I_z$ are the intermediate variables introduced in \cite{rys_first, rys_quadrature}.  For every root, each of the intermediate variables is a tensor of shape $(l_i+1,l_j+1,l_k+1,l_l+1)$.  The total number of intermediate variables is $3\times 3 \times (2 \times 2 \times 2 \times 2) = 144$ for calculating a $(pp|pp)$ integral (the first 3 refers to the three Cartesian dimensions, and the second 3 is the number of roots). For $(gg|gg)$ integral, the number of variables is raised up to $3\times 9 \times (5\times 5\times 5\times 5) = 16875$, which exceeds the capacity of GPU registers. The size of intermediate variables can be reduced if the roots are handled sequentially. The same space can be reused for different roots. The intermediate variables $I_x, I_y, \text{and } I_z$ depend on the exponents, atomic coordinates and angular momentum of the four primitive Gaussians only. They can be efficiently evaluated using recursive relations, documented in \cite{rys_quadrature}. But storing those values on the limited resources on GPU can be tricky, and several strategies will be discussed in Sec. \ref{sec:1t1q} and Sec. \ref{sec:1qnt}. The index set for enumerating the Gaussian functions of a primitive shell is defined as
$$S_l := \{(i,a,b,c) | a+b+c=l, i=\texttt{idx}(a,b,c)\}$$
The \texttt{idx} function is defined differently in various quantum chemistry packages.  We adopt reversed lexicographic order: for $l = 2$ (d shell), $\texttt{idx}(2,0,0) = 0$, $\texttt{idx}(1,1,0) = 1$, ..., $\texttt{idx}(0,0,2) = 5$. Note that other packages may adopt different
  orderings; the density matrix and orbital coefficients must be reordered accordingly. 
  The indexing operation in general makes the memory access random and less efficient.

Because of its huge size, the four-dimensional integral tensor is not formed explicitly in the main memory, but instead contracted with the density matrix $D$ to form the Coulomb ($J$) and exchange ($K$) potential matrices:
\begin{equation}
\label{eq:jk}
    J_{ij} = \sum_{kl}(ij|kl)D_{kl}, \text{ and } K_{ik} = \sum_{jl}(ij|kl)D_{jl}
\end{equation}
For a molecular system with real-valued orbitals, the 8-fold symmetry of $(ij|kl)$ can be used to save computational cost,
\begin{equation}
(ij|kl)
=
(ji|kl)
=
(ij|lk)
=
(ji|lk)
=
(kl|ij)
=
(lk|ij)
=
(kl|ji)
=
(lk|ji).
\end{equation}

 As a result, only integrals $(ij|kl)$ where $i > j$, $k > l$ and $i N_{ao} + j > k N_{ao} + l$ needs to be evaluated, and this saves $7/8$ of the integral evaluation work. In particular, for the integral element $(ij|kl)$ composed of the $i$-th, $j$-th, $k$-th and $l$-th atomic orbital,  the J matrix requires a trace over the bra-ket pair $(k,l)$, summing $(ij|kl)D_{kl}$, yielding two updates ($J_{ij}$ and $J_{ji}$ by symmetry). The K matrix involves cross-contractions where one index from the bra pairs and one from the ket, yielding four distinct updates. Therefore, the integral $(ij|kl)$ is contracted with two density matrix elements $D_{kl}$ and $D_{lk}$ to form two matrix elements $J_{ij}$ and $J_{ji}$. And it is simultaneously contracted with four density matrix elements $D_{jl}$, $D_{jk}$. $D_{il}$, $D_{ik}$ to form four $K$ matrix elements $K_{ik}$, $K_{il}$, $K_{jk}$, $K_{jl}$.

Conventionally, the Schwarz inequality for shell quartets
\begin{equation}
\label{eq:schwartz}
|(ij|kl)|
\;\le\;
\sqrt{(ij|ij)}\sqrt{(kl|kl)}.
\end{equation}

is used to screen the significant shell quartets, i.e. all integrals within an insignificant shell quartet are considered negligible and not evaluated. The cost of generating the significant shell quartet is almost negligible, even for the low-angular-momentum integrals. Other screening techniques can be combined with the algorithms. They will be discussed in section \ref{sec:future}.

In the direct self-consistent field (SCF) algorithm, the integrals are further screened by the density matrix
\begin{equation}
    |\Delta J_{ij}| \le \sqrt{(ij|ij)}\sum_{kl}| \Delta D_{kl}|\sqrt{(kl|kl)}
\end{equation}
and
\begin{equation}
    |\Delta K_{ik}| \le \sum_{jl} |\Delta D_{jl}|\sqrt{(ij|ij)}\sqrt{(kl|kl)}
\end{equation}
where $\Delta D$, $\Delta J$, and $\Delta K$ are the changes of density matrix, Coulomb potential, and exchange potential between two steps of SCF iterations (i.e. incremental SCF), respectively. Since $\Delta D$ decreases along the SCF iterations, the screening is more effective. In practice, the density matrix screening is performed at shell pair level, which is similar to Eq. \eqref{eq:schwartz}.
\section{Design and Algorithms}
The central design decision in a JIT framework is which inputs to treat as compile-time constants (static) and which to pass as runtime arguments (dynamic). Making too few variables static forfeits optimization opportunities; making too many static causes excessive recompilation. We adopt the following rule: protocol and device information — the quantum chemistry method, basis set type, and GPU architecture — are static, while molecular information — atomic coordinates, Gaussian orbital exponents, and contraction coefficients — remains dynamic. This ensures that a single compilation serves all molecules sharing the same basis set, while still enabling aggressive loop unrolling and register allocation.

More specifically, we select the angular momentum and contraction pattern of the basis set, as well as the task type, as the template parameters for the kernels. They are described in Table \ref{tab:template_params}. 
 A crucial consequence of this design: once the angular momentum, contraction pattern, and number of primitives are fixed, every loop bound and array size in the integral kernel
  becomes a compile-time constant. The compiler can then fully unroll loops, allocate registers statically, and eliminate all branching.
If those parameters are changed, we have to recompile the kernels. With such choices, we avoid complicated runtime logic flows for different data types, task types, and unrolling strategies. Thus the amount of source code is minimal: only \textasciitilde1,000 lines of code for each algorithm, much less than the \textasciitilde20,000 lines in GPU4PySCF v1.4.

\begin{table}[h]
    \centering
    \begin{tabular}{c|c|l}
        \makecell{Template \\Parameter} & Data Type & Description \\
        \hline
        $(l_i, l_j, l_k, l_l)$ & int     & Angular momentum of shells \\
        \texttt{do\_j}         & bool    & Whether to compute the J matrix \\
        \texttt{do\_k}         & bool    & Whether to compute the K matrix \\
        \texttt{rys\_type}      & int     & 0: no range separation (full range), \\
                                &        & 1: long-range, -1: short-range \\
        \texttt{n\_dm}         & int     & \# of density matrices \\
                                &       & (e.g. two for unrestricted DFT) \\
        \texttt{DataType}      & type    & \texttt{float} or \texttt{double} precision \\
        $(n_i^p, n_j^p, n_k^p, n_l^p)$ & int     & \# of primitives in the \\
                             &      & (i, j, k, l) shells \\
        $(F_i, F_j,F_k, F_l)$ & int     & Fragment sizes of the \\
                             &      & $(i, j, k, l)$ shells (details in Sec. \ref{sec:1qnt})\\
    \end{tabular}
    \caption{Static parameters for integral kernel generation.}
    \label{tab:template_params}
\end{table}

The parameters of the basis set are dynamic variables. Since many elements share the same contraction pattern but different coefficients and exponents, we do not need to compile multiple kernels for the same contraction pattern. 
The basis set parameters (exponents, coefficients, coordinates) are stored in global memory and read once per shell quartet. Since the data size is usually small and is efficiently cached through the L1/L2 cache, reading those data is not the bottleneck of the kernels. Another strategy is to form the basis pair data as the kernel input, and the pair data can be more efficient in exp operation in some scenarios, for example F64 kernel on RTX cards, at the cost of less efficient cache utilization.  Directly taking the basis data as input is part of the current design, but may not be optimal.

All of the dynamic parameters are summarized in \ref{tab:input_params}.
\begin{table}[h]
    \centering
    \begin{tabular}{c|c|l}
        \makecell{Input} & Data Type & Description \\
        \hline
        $(c_i, c_j, c_k, c_l)$ & double/single  & contraction coefficients in basis \\
        $(e_i, e_j, e_k, e_l)$ & double/single  & exponents in basis \\
        \texttt{coords}        & double/single  & coordinates of atoms \\
        $D_{ij}$         & double     & density matrix as input \\
        $J_{ij}$          & double     & coulomb potential as output \\
        $K_{ij}$          & double     & exchange potential as output \\
        $N_{ao}$          & int        & \# of atomic orbitals \\
        $\omega$          & double     & range-separation parameter \\
    \end{tabular}
    \caption{Dynamic parameters as the input of integral kernels. The data type of basis parameters are determined by the static variable DataType.}
    \label{tab:input_params}
\end{table}

\begin{figure}
    \centering
    \includegraphics[width=0.45\linewidth]{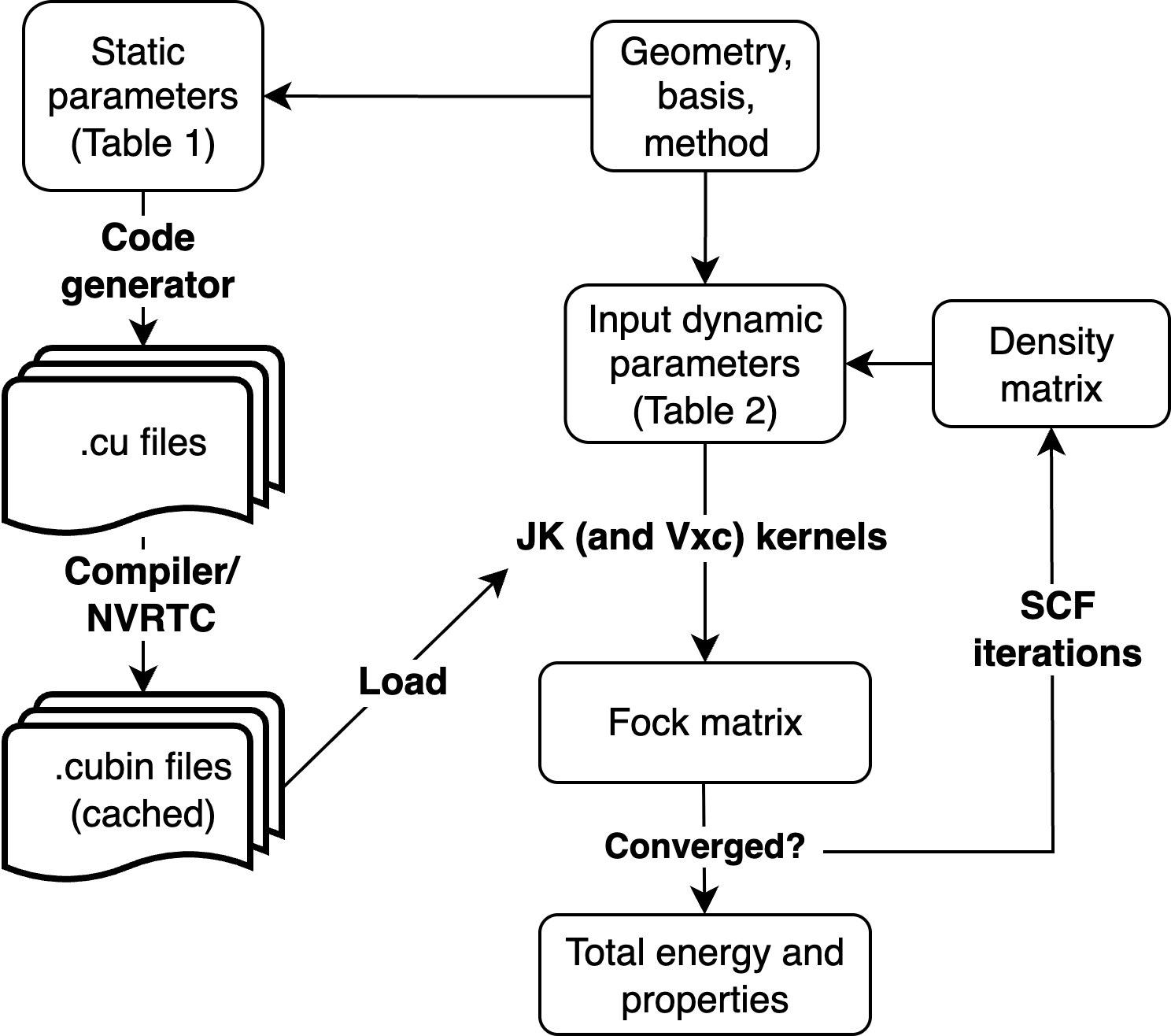}
    \caption{Workflow of SCF iterations with NVRTC as JIT compiler.}
    \label{fig:diagram}
\end{figure}

An example of the DFT self-consistent field (SCF) workflow is presented in Fig. \ref{fig:diagram}. The workflow employs NVRTC \cite{nvrtc}, the runtime compilation library for CUDA, via CuPy\cite{cupy_learningsys2017} as JIT compiler. NVRTC offers the direct control over shared memory, as well as a relatively fast compilation. Other compilers, such as Numba, can be employed similarly with minimal modification, which will be discussed in Sec. \ref{sec:future}

Based on this design, we build heuristics (including algorithms and fragmentation strategies) to deliver optimized kernels for a specific device type and precision type (FP64/FP32). The heuristics are benchmarked ahead of time, and the best ones are saved in a database as a part of the final package. In the runtime, the kernel caller will find the corresponding algorithm based on the device type and required precision. Then, only one optimized kernel is compiled. It should be noted that, even for the exact same integrals, with different device/precision combination, different heuristics might be selected. In particular, the FP64 algorithm may be different from the FP32 algorithm for the same angular momentum, in order to achieve the best performance. 

In the next two sections, we present two heuristics algorithms for integral kernels, each targeting a different angular-momentum regime. The 1q1t algorithm (Sec. \ref{sec:1t1q}) assigns one thread per shell quartet and is often well-suited for low angular momentum (s, p shells), where the integrals fit in registers. The 1qnt algorithm (Sec. \ref{sec:1qnt}) fragments the integral across multiple threads and is designed for high angular momentum
  (d, f, g shells), where register pressure becomes the bottleneck.

\subsection{One quartet one thread (1q1t) algorithm} \label{sec:1t1q}
The first algorithm we present is largely the same as the classical Rys quadrature algorithm on CPUs \cite{libcint}, except that a more aggressive loop unrolling strategy is used. First, a list of significant shell quartets is generated by running over all possible combinations screened by the Schwartz inequality. Each quartet is assigned to one thread for evaluating all elements of the integral in Eq. \eqref{eq:rys0} (including all primitives in a contraction and all functions in a shell), and contract it with the density matrix according to Eq. \eqref{eq:jk}. This algorithm was implemented in the early version of GPU4PySCF (v1.0 and before). This straightforward implementation has two performance issues. 1) Several performance-critical loops — over primitives in each shell and over density matrices — have bounds unknown at compile time and therefore cannot be unrolled. On GPUs, unrolled loops enable the compiler to schedule instructions more efficiently, eliminate branch overhead, and allocate registers to specific loop iterations. This is the dominant bottleneck for low-angular-momentum integrals, which are individually cheap but numerous. JIT compilation resolves this by making all loop bounds compile-time constants. 2) Significant register spills to local memory for high angular momentum. In this subsection, we show that with more known variables at compilation time, the performance is significantly improved. The second issue will be discussed in the next subsection.

With the design in this work, when handling one shell quartet, the ranges of all loops are known at compilation time. The JIT compiler can aggressively unroll those loops, especially loops for primitives, and use registers more efficiently.

In the single-precision implementation, the exponents, contraction coefficients, and atomic coordinates are prepared and loaded in FP32. 
The density matrix is maintained in FP64 throughout the SCF procedure for diagonalization and DIIS, so we load it in its native format and cast to FP32 inside the kernel to avoid maintaining a separate single-precision copy. One may improve the memory throughput by maintaining a separate single-precision copy.
Inside the kernel, all the arithmetic operations are performed in FP32 and accelerated with the \texttt{{-}{-}use\_fast\_math} compiler flag. 
The \texttt{{-}{-}use\_fast\_math} flag replaces standard math functions (\texttt{sqrt}, \texttt{exp}, \texttt{erf}) with hardware-accelerated intrinsics that are faster but not IEEE-754 compliant (i.e. they introduce an additional small rounding error). The accelerated \texttt{exp} function benefits the evaluations of Gaussian functions, as the kernel takes the native Gaussian exponents and contraction coefficients as inputs. The pair data is calculated inside the kernel. The benefit of \texttt{exp} is negligible if one precompute the pair data before invoking the JK kernel. The calculations of Rys roots and weights are accelerated as well for two reasons. 1) We directly use \texttt{erf} to evaluate the root and weight when $N_r=1$ (interpolation is used for other cases), as we found that directly evaluating the special functions is often faster than the interpolation. 2) In the long-distance regime for $N_r>1$ (outside of interpolation window) of Rys roots and weights, \texttt{sqrt} is used in the asymptotic formula. \cite{purple_book}

\begin{figure}[htbp]
    \centering
    \includegraphics[width=0.7\textwidth]{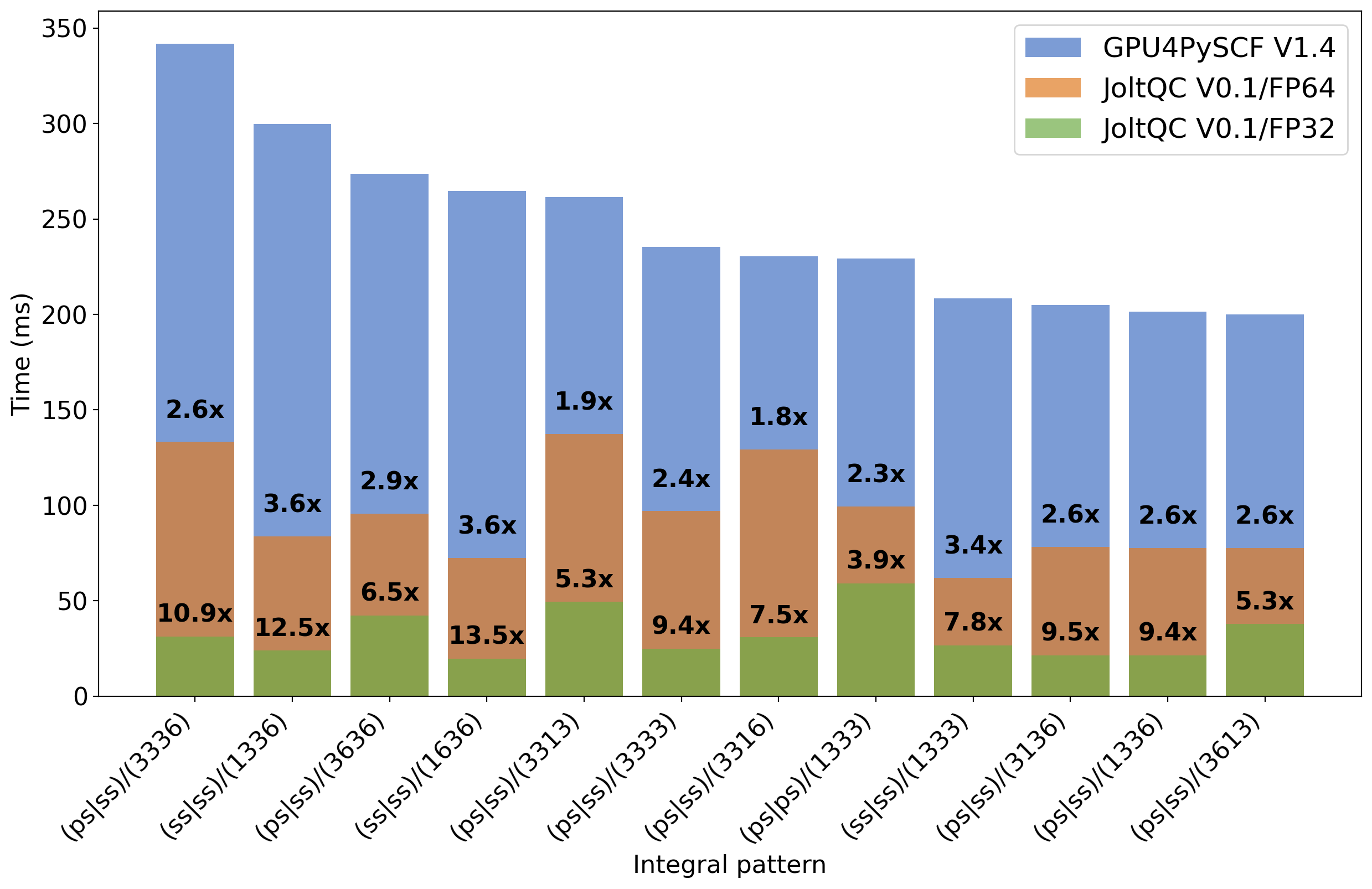}
    \caption{Averaged kernel runtime of JoltQC v0.1 JK calculation using 1q1t algorithm, on the gly120 system, 6-31G basis set. One NVIDIA A100-80G GPU is used. The performance is compared with GPU4PySCF v1.4. Top 12 most time-consuming kernels are selected, and are labeled as (angular momentum combination)/(contraction pattern combination).}
    \label{fig:6-31g_A100}
\end{figure}

\begin{figure}[htbp]
    \centering
    \includegraphics[width=0.7\textwidth]{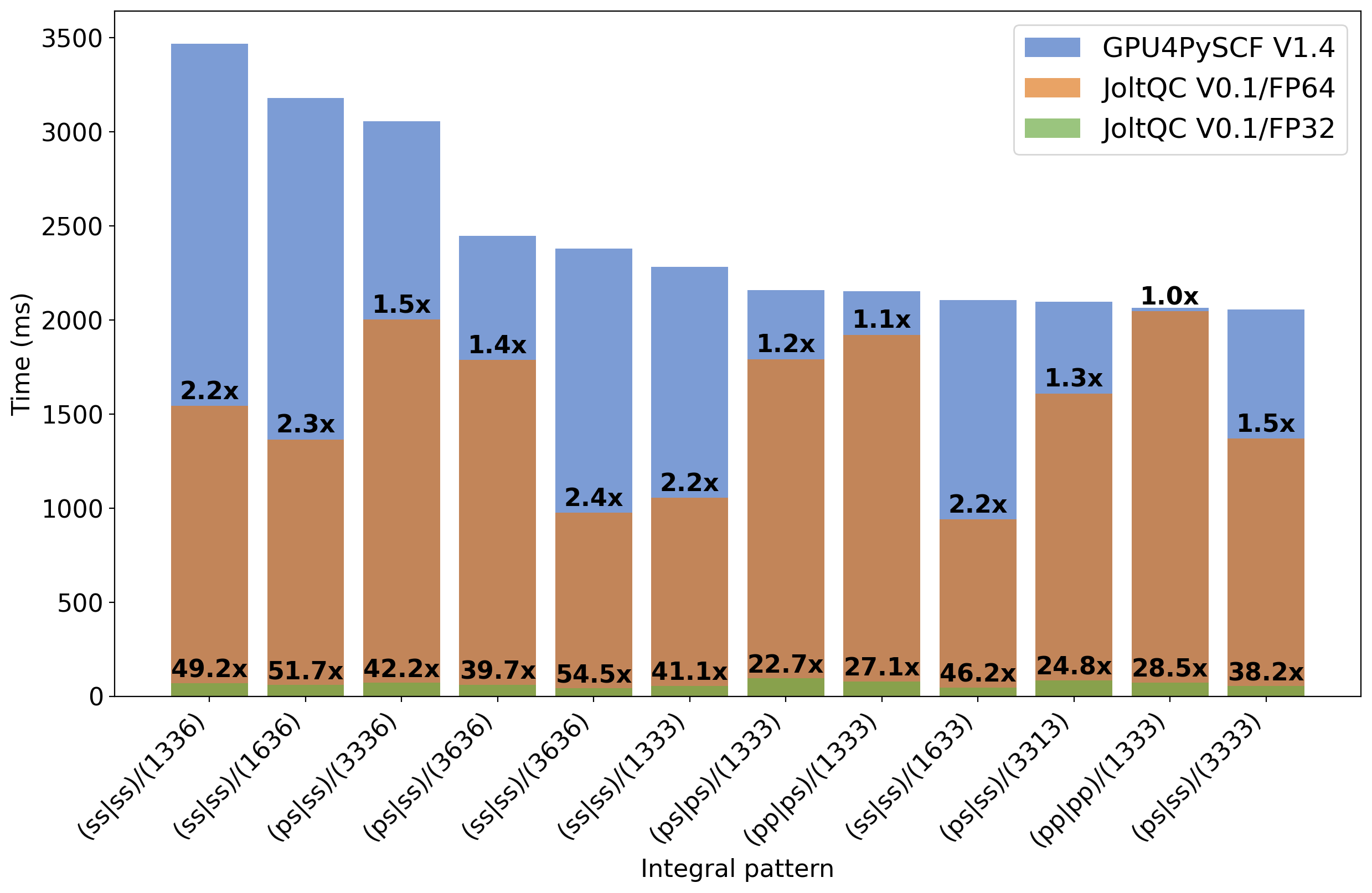}
    \caption{Averaged kernel runtime of JoltQC v0.1 JK calculation using 1q1t algorithm, on the gly120 system, 6-31G basis set. One NVIDIA A10-24G GPU is used. The performance is compared with GPU4PySCF v1.4. Top 12 most time-consuming kernels are selected, and are labeled as (angular momentum combination)/(contraction pattern combination).}
    \label{fig:6-31g_A10}
\end{figure}

We benchmark the performance of JK kernels with a straight peptide chain composed of 120 glycines (gly120), and 6-31G \cite{6-31g} basis set. Density matrix screening is turned off for profiling purpose. The timings on A100-80G and A10-24G are shown in Figure \ref{fig:6-31g_A100} and \ref{fig:6-31g_A10} respectively. The overall speedup of FP64 against AOH compilation in GPU4PySCF v1.4 is about 2.0x. For low-angular-momentum, the evaluations of \texttt{exp} and \texttt{erf} functions are often the computational bottleneck. Thanks to the special hardware acceleration of these functions on GPU, the overall speedup of FP32 reaches about 5-10x on A100-80G. 
 Since the A10-24G only delivers 1/32 of P64 throughput compared to FP32,
FP64 algorithms are always bounded by the limited compute units. The improvement with JIT is marginal. When switched to FP32, the 1q1t algorithm is accelerated by 30x.

There are still several corner-case kernels. When the number of primitives is greater than 5, the NVRTC compiler refuses to unroll the nested loops, even if we specify \texttt{\#pragma unroll}. Performance drop is observed. The issue will be discussed in Appendix (Section \ref{sec:appendix_analyze}). Another case is $(pp|pp)$ in double precision which requires 96 registers per thread for the intermediate variables, and 162 registers for the integral elements. The kernel consumes at least 258 registers, which already exceeds the hard limit per thread (255 registers) by the CUDA platform. The register spill affects the performance dramatically. Higher angular momentum integrals suffer from the same issues. They will be handled by different algorithm later in Section \ref{sec:1qnt}.

\subsection{Quartet fragmentation (1qnt) algorithm}
\label{sec:1qnt}
For high-angular-momentum integrals, the intermediate variables and integral elements of one shell quartet cannot fit in the registers. From GPU4PySCF v1.0-v1.4, a fixed-size 1D fragmentation strategy was introduced to split the integral into multiple threads. Instead of a single thread processing all integrals of one shell quartet, a group of 2 to 256 threads cooperatively compute all integral elements within one shell quartet. The strategy significantly improved the performance of the high-angular-momentum integrals. In this work, we introduce the dynamic fragmentation strategy to further improve the data locality, and multi-level reduction algorithm to reduce the bottleneck of memory access (both shared memory and global memory). Since multiple threads for the same shell quartet use the same intermediate variables, the intermediate variables $I_x$, $I_y$ and $I_z$ are stored in shared memory. Then, each thread in the group will compute and store into registers only a fraction of the integrals. For example, for $(dd|dd)$, the integral tensor has shape (6,6,6,6). With fragment size (3,3,3,3), each dimension is split into 6/3 = 2 parts. The four fragment indices ($f_i$, $f_j$, $f_k$, $f_l$) are distributed across threads, requiring $2^4 = 16$ threads per shell quartet. 
 After computing its integral fragment $T^{reg}$, each thread contracts $T^{reg}$ with the density matrix to produce local contributions to J and K. We denote the thread-local J contribution as $v_j$ (shape $F_i \times F_j$) and the thread-local K contribution as $v_k$ (shape $F_i \times F_k$). These are accumulated in registers during the thread-level reduction (Algorithm \ref{fragment_algorithm}, lines 21–30), then reduced across threads at the block level (lines 32–36).

The reduction scheme is presented in Algorithm \ref{fragment_algorithm}. The shape of the intermediate variables are also known at compilation time. A similar idea has been proposed in \cite{1qnt_first}, where $(dd|dd)$ is partitioned in $i$, $j$, and $k$ indices, while $(ff|ff)$ is partitioned in $i$ and $j$ indices.

\begin{algorithm}
\caption{1qnt multi-level reduction algorithm (without 8-fold symmetry)}
{
\setlength{\baselineskip}{0.6\baselineskip}
\begin{algorithmic}[1]
\REQUIRE Input value shell quartet id, basis info
\ENSURE Output value $J$ and $K$ matrices
\STATE // Superscripts: reg = per-thread registers, shd = per-block shared memory, no superscript = global memory.
\STATE $J^\text{shd}[\colon\!N^f_i, \colon\!N^f_j] = 0$, $K^\text{shd}[\colon\!N^f_i, \colon\!N^f_k] = 0$ \hfill // Allocate J, K  in shared memory
\STATE //{\it Distribute fragments to different threads}
\FOR{$f_i = 0$ to $N^f_i/F_i$ in parallel}
\FOR{$f_j = 0$ to $N^f_j/F_j$ in parallel}
\FOR{$f_k = 0$ to $N^f_k/F_k$ in parallel}
\FOR{$f_l = 0$ to $N^f_l/F_l$ in parallel}
\STATE // Integral fragment $T_{ijkl} = (ij|kl)$
\STATE $T^{\text{reg}}[\colon\!N^f_i, \colon\!N^f_j, \colon\!N^f_k, \colon\!N^f_l] = 0$ \hfill // Allocate integral fragment in registers
\FOR{$i_p = 0$ to $n_i^p$}
\FOR{$j_p = 0$ to $n_j^p$}
\FOR{$k_p = 0$ to $n_k^p$}
\FOR{$l_p = 0$ to $n_l^p$}
\STATE // {\it Compute integrals with Eq. \eqref{eq:rys0}-\eqref{eq:rys2}}
\STATE $T^{\text{reg}}[i,j,k,l] \mathrel{+}= c_{ip} c_{jq} c_{kr} c_{ls}(pq|rs)$
\ENDFOR
\ENDFOR
\ENDFOR
\ENDFOR
\STATE Read $D_J^{\text{reg}}[:\!F_k,:\!F_l]$ and $D_K^{\text{reg}}[:\!F_j,:\!F_l]$ \hfill  // Read density matrix subblocks in registers
\STATE $J^\text{reg}[\colon\!F_i, \colon\!F_j]=0, K^\text{reg}[\colon\!F_i,\colon\!F_k]=0$ \hfill  // Allocate accumulators for J, K in registers
\FOR{$i = 0$ to $F_i$}
\FOR{$j = 0$ to $F_j$}
\FOR{$k = 0$ to $F_k$}
\FOR{$l = 0$ to $F_l$}
    \STATE //{\it Thread level reduction}
    \STATE $J^{reg}[i,j] \mathrel{+}= \sum_{kl}T^{\text{reg}}[i,j,k,l]D_{J}^{\text{reg}}[k,l]$
    \STATE $K^{reg}[i,k] \mathrel{+}= \sum_{jl}T^{\text{reg}}[i,j,k,l]D_{K}^{\text{reg}}[j,l]$
\ENDFOR
\ENDFOR
\ENDFOR
\ENDFOR
\STATE //{\it Block level reduction}
\STATE $J^\text{shd}[iF_i\!:\!(i+1)F_i,\,jF_j\!:\!(j+1)F_j] \mathrel{+}= J^\text{reg}[\!:\!F_i,\!:\!F_j]$
\STATE $K^\text{shd}[iF_i\!:\!(i+1)F_i,\,kF_k\!:\!(k+1)F_k] \mathrel{+}= K^\text{reg}[\!:\!F_i,\!:\!F_k]$
\ENDFOR
\ENDFOR
\ENDFOR
\ENDFOR
\STATE atomicAdd($J[:\!N^f_i,:\!N^f_j], J^\text{shd}[:\!N^f_i,:\!N^f_j]$) \hfill  // Write block results to global J
\STATE atomicAdd($K[:\!N^f_i,:\!N^f_k], K^\text{shd}[:\!N^f_i,:\!N^f_k]$) \hfill  // Write block results to global K
\end{algorithmic}
}
\label{fragment_algorithm}
\end{algorithm}

{\bf Optimize the fragmentation scheme}. The performance of the integral kernel is sensitive to the fragment size. Different fragmentation strategies can easily lead to a 10x difference in performance. The fragment size can impact the register usage of each thread, the shared memory usage of each thread block, and the number of shell quartet assigned to each block. The resource consumption also need to meet the hard requirements of CUDA compute capability. We have to take these factors into account for searching the optimal fragment size. This optimization problem is unlikely to be solved analytically. We look for the optimal fragmentation strategy by grid search. For each candidate fragmentation, we compile and execute the kernel on a representative workload and select the configuration with the lowest measured runtime. The searching space of high angular momentum can be huge. We made several assumptions:
\begin{itemize}
\item We assume that each dimension of the integral is a multiple of the fragment size. There probably exists a better tiling strategy without this constraint. However, based on the grid search results, we found that the divisible fragment size, in general, is optimal. An additional benefit of this constraint is that each thread does not need to check the bound of iteration, which saves the significant computational cost.

\item The optimal fragment size is independent of the number of primitives. This assumption in general is not true, since the kernel with more number of primitives take more registers. Then, less registers are left for the integral fragment. 
This assumption is highly empirical. In actual runs, kernels with more primitives consume more registers, leaving fewer for the integral fragment.
However, jointly optimizing over both fragment size and primitive count would make the search space intractable. We therefore fix the primitive count at a representative value and search only over the space of fragment size. For high-angular-momentum shells ($d$, $f$, $g$), which typically have fewer primitives, we set the primitive count to 1.

\item The block size is always 256 threads.
\end{itemize}

 We optimize fragmentation strategies separately for the combinations of precisions and GPU devices. As previously mentioned, the FP64 algorithm may use a different fragmentation scheme from the FP32 algorithm for the same angular momentum. Since the double-precision and single-precision algorithms differ in their usage of shared memory, register count, and hardware units, a strategy optimized for double precision is not necessarily optimal for single precision. The default fragmentation schemes are tuned on NVIDIA A100-80G and A10-24G GPUs, respectively. To ensure compatibility, the fragment size is optimized under the constraint of 48 KB shared memory per streaming multiprocessor, allowing the strategy to be applicable to older GPU generations. An optimization script is included in the JoltQC release, enabling users to retune the fragment size for different GPU architectures if they diverge significantly from the default targets.

\begin{figure}[htbp]
    \centering
    \includegraphics[width=0.7\textwidth]{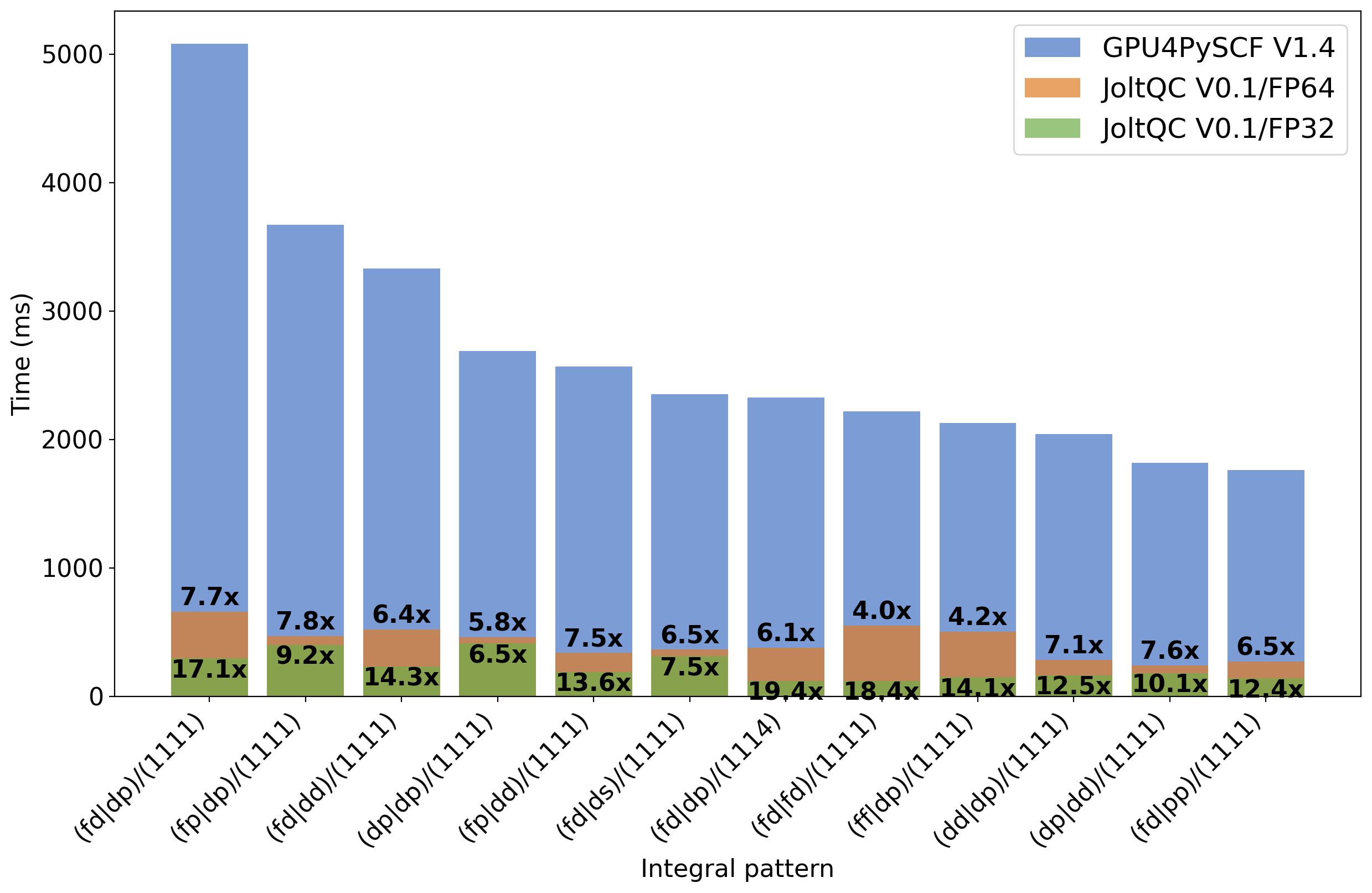}
    \caption{Averaged kernel runtime of JoltQC v0.1 JK calculation using 1qnt algorithm, on the gly30 system, def2-TZVPP basis set. One NVIDIA A100-80G GPU is used. The performance is compared with GPU4PySCF v1.4. Top 12 most time-consuming kernels are selected, and are labeled as (angular momentum combination)/(contraction pattern combination).}
    \label{fig:tzvpp_a100}
\end{figure}

\begin{figure}[htbp]
    \centering
    \includegraphics[width=0.7\textwidth]{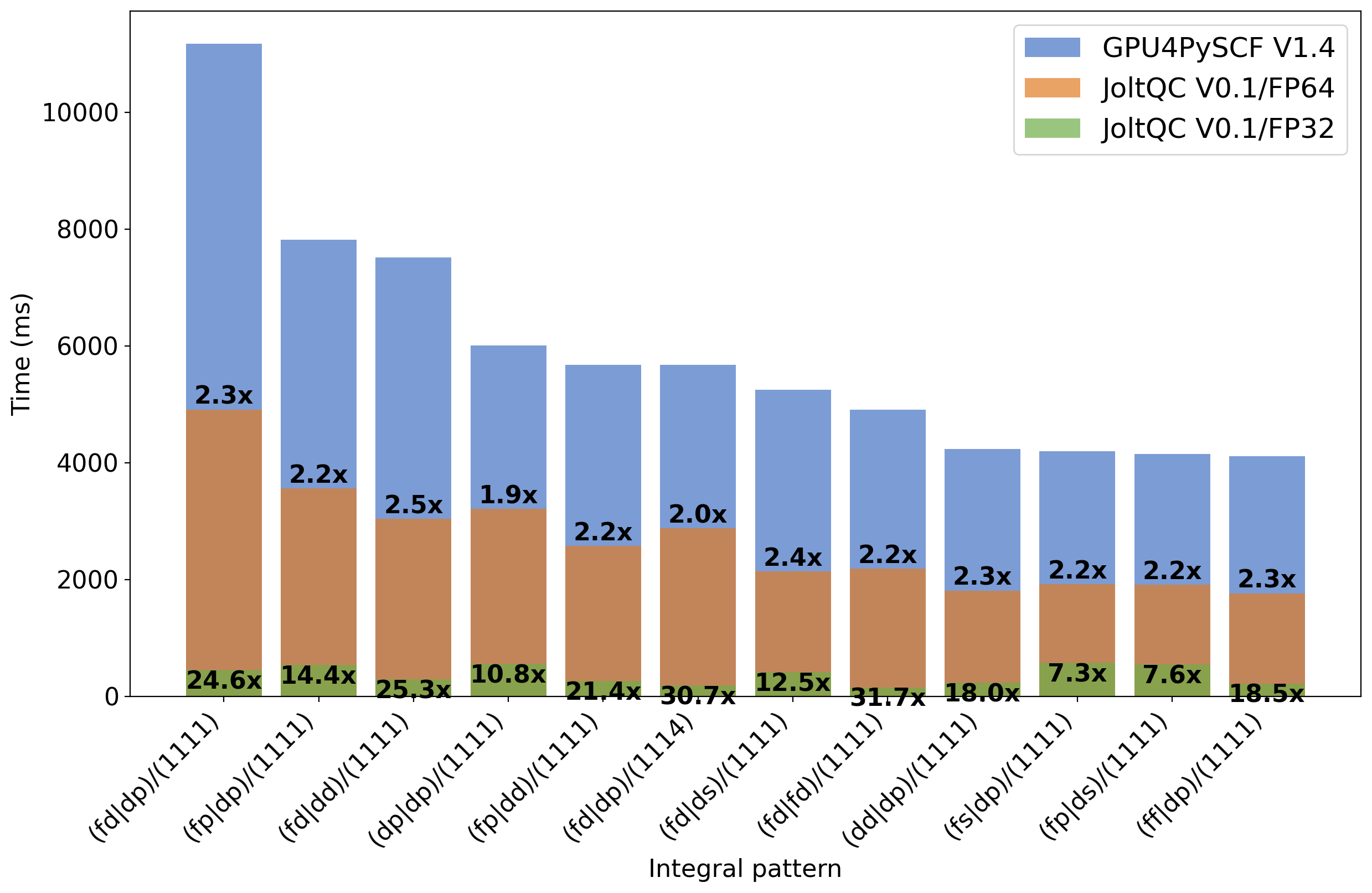}
    \caption{Averaged kernel runtime of JoltQC v0.1 JK calculation using 1qnt algorithm, on the gly30 system, def2-TZVPP basis set. One NVIDIA A10-24G GPU is used. The performance is compared with GPU4PySCF v1.4. Top 12 most time-consuming kernels are selected, and are labeled as (angular momentum combination)/(contraction pattern combination).}
    \label{fig:tzvpp_a10}
\end{figure}

We benchmark the 1qnt algorithm with a straight peptide chain composed of 30 glycines (gly30) and def2-TZVPP \cite{def2-tzvx} basis set. Density matrix screening is turned off. Results on NVIDIA A100-80G and A10-24G GPUs appear in Figure \ref{fig:tzvpp_a100} and \ref{fig:tzvpp_a10}, respectively. Using the default fragmentation for FP64, the 1qnt algorithm is 5-8x faster than GPU4PySCF v1.4 on A100-80G. When switched to FP32, the algorithm is further accelerated by 2x. For high-angular-momentum integrals, the computational cost of \texttt{exp} and \texttt{erf} functions is not significant relative to multiply/add operations. 

On A10-24G, the high-angular-momentum integrals are accelerated by roughly 2x for FP64. This is due to the algorithm in GPU4PySCF v1.4 being memory bound. The 1qnt algorithm reduces the memory bottleneck by multi-level reduction. When switched to FP32, the integrals are accelerated by 10-20x. Further performance analysis is provided in Appendix (Section \ref{sec:appendix_analyze}).

\section{Results}
We evaluate JoltQC along three axes: compilation overhead (Sec. \ref{sec:overhead}), end-to-end SCF performance against GPU4PySCF v1.4 and TeraChem v1.9 (Sec. \ref{sec:benchmark}), and single-precision accuracy. All benchmarks use a single NVIDIA A100-80G GPU for FP64 and an A10-24G GPU for FP32 unless otherwise noted.
\subsection{Compilation Overhead}
\label{sec:overhead}
Just-in-time compilation introduces non-negligible overhead, but this cost is amortized through on-disk caching and in-memory reuse. When a kernel is first invoked, it is compiled and the resulting binary is saved to disk. The cached binaries can be reused across the processes. During the first SCF iteration, each process loads these binaries into RAM directly; all subsequent iterations then fetch them directly from memory. Table \ref{tab:widgets} summarizes the one-time compilation times for various kernels using CUDA v12.1. As expected, high–angular-momentum integrals take longer to compile and generate larger binaries: for the C, H, O, and N elements with the 6-31G* basis (140 kernels), the compilation time is just 30 s and the total kernel size is roughly 10 MB, while the def2-TZVPP basis (750 kernels) yields about 100 MB of binary and takes approximately 200 s to compile. Once those kernels are compiled, subsequent calls take less than 1 s to load cached binaries.

These figures serve as a reference. Actual compilation times will vary with different CUDA versions and hardware quality, particularly disk performance. In practice, the compilation overhead can be reduced in various ways. Especially in homogeneous cloud environments, it can be advantageous to store compiled kernels on shared or remote storage and download them as needed. 
  A similar approach is used in the machine learning community, where PyTorch's AOTInductor pre-compiles GPU kernels and caches the resulting binaries for reuse across runs. Alternatively, retaining the intermediate PTX assembly (a CUDA specific compilation intermediate) and loading it directly allows users to skip the final machine-code generation step on subsequent invocations.
\begin{table}
\centering
\begin{tabular}{l|c|r|r}
Integral pattern & Algorithm & \makecell{Compilation\\time (ms)} & \makecell{Binary size\\(KB)} \\\hline
$(ss|ss)/(11|11)$ & 1q1t & 284.8 & 25\\
$(ss|ss)/(33|33)$ & 1q1t & 512.0 & 74\\
$(pp|pp)/(11|11)$ & 1q1t & 638.8 & 75\\
$(pp|pp)/(33|33)$ & 1q1t & 880.6 & 94\\
$(dd|dd)/(11|11)$ & 1qnt & 6,011 & 194\\
$(ff|ff)/(11|11)$ & 1qnt & 7,697 & 270\\
$(gg|gg)/(11|11)$ & 1qnt & 59,468 & 602\\
\end{tabular}
\caption{\label{tab:widgets}Compilation cost of selected integral kernels with NVRTC in CUDA V12.1. All of the kernels are compiled in FP64.}
\end{table}

\subsection{Benchmark with GPU4PySCF and TeraChem}
\label{sec:benchmark}
In this work, integrals for the Coulomb (J) and exchange (K) terms are screened simultaneously. GPU4PySCF and JoltQC use the same screening technique. The number of significant shell quartets in each JK calculation is identical in both cases. 
In double precision, JoltQC reproduces GPU4PySCF's J and K matrices to machine precision for all test systems, confirming that the JIT-compiled kernels are numerically identical to the AOT implementation. We therefore focus the accuracy analysis on the single-precision results.
Under our single precision scheme, JK matrix elements are computed in single precision but accumulated in FP64; all other operations including diagonalization and direct inversion of the iterative subspace (DIIS)—remain in FP64.

We use a customized TeraChem v1.9 build with f-orbital support. 
 Before presenting timings, we note several differences that affect the fairness of comparisons. GPU4PySCF and JoltQC use identical screening (Schwarz inequality at the contracted level), so their JK kernel timings are directly comparable. TeraChem applies Schwarz screening at the primitive shell level, uses different SCF initial guesses and DIIS settings, and converges in fewer iterations (e.g. 11 vs. 14 for the gly30 molecule). Consequently, GPU4PySCF/JoltQC–TeraChem comparisons reflect differences in both algorithms and settings, and are not strictly apples-to-apples. For all calculations, the screening threshold is set to $10^{-13}$.  SCF with GPU4PySCF/JoltQC converges in 14 iterations, whereas the same SCF with TeraChem converges in 11 iterations. This does not imply TeraChem's settings are superior, as convergence behavior can vary with the system. A sample TeraChem input is provided in Appendix \ref{sec:terachem_input}. All the basis sets are in Cartesian format.

When both codes run in FP64, their total energies agree to within 0.001 mHa. When all integrals are evaluated in FP32, we observe uncertainties on the order of 1 mHa in both JoltQC and TeraChem (Table \ref{tab:accuracy}), reinforcing the well-known result that pure single-precision arithmetic is insufficient for high-accuracy quantum-chemical calculations. For the cc-pVQZ basis, which involves higher angular momentum, the FP32 energy deviations remain sub-milliHartree (0.16 mHa for tamoxifen and 0.23 mHa for sphingomyelin), indicating that single-precision accuracy does not degrade with increasing basis set size. A more practical mixed-precision algorithm is among our future directions.

\begin{table*}
    \centering
    \begin{tabular}{l|l|r|r|r}
       Molecule & Basis & \makecell[r] { GPU4PySCF/FP64 \\ - JoltQC/FP32 } & \makecell[r]{GPU4PySCF/FP64 \\ - TeraChem/FP64} & \makecell[r] {JoltQC/FP32 \\- TeraChem/FP32}\\
       \hline
       gly30  &  6-31G*  &  0.23 mHa & 0.0005 mHa & -0.07 mHa\\
       gly30  &  def2-TZVPP & 1.95 mHa & 0.0004 mHa & 2.10 mHa\\
       valinomycin & 6-31G* &  -0.06 mHa & -0.001 mHa & -0.26 mHa \\
       valinomycin & def2-TZVPP & 0.68 mHa & -0.001 mHa & 0.63 mHa\\
       Tamoxifen & cc-pVQZ &  0.16 mHa & - & -  \\
       Sphingomyelin & cc-pVQZ &  0.23 mHa & - & -  \\
    \end{tabular}
    \caption{\label{tab:accuracy} The accuracy of Hartree-Fock (HF) energy with JoltQC v0.1 against TeraChem v1.9. GPU4PySCF/FP64 and JoltQC/FP64 matches exactly and is not shown in the table. Dashes indicate that the software does not support the required angular momentum (g functions in cc-pVQZ). Each column reports the total energy difference $E_A - E_B$ between the two codes listed.}
\end{table*}

As shown in Table \ref{tab:benchmark_fp64}, JoltQC delivers a twofold speedup over GPU4PySCF v1.4 when using the 6-31G* basis set and a fourfold gain with the larger def2-TZVPP basis. Although TeraChem v1.9 generally outpaces GPU4PySCF v1.4, its advantage is only marginal. In double-precision, JoltQC outperforms TeraChem by about 4x when the def2-TZVPP basis set is used, and a less significant 1.6x for the 6-31G* basis. The advantage grows further with larger basis sets. For the cc-pVQZ basis, which includes g-type functions not supported by TeraChem v1.9, JoltQC achieves a 7–8× speedup
  over GPU4PySCF v1.4 on the tamoxifen, raffinose, and sphingomyelin systems — nearly doubling the gain observed with def2-TZVPP. This trend confirms that JIT compilation becomes increasingly beneficial as the angular momentum of the basis set rises, precisely the regime where register pressure and loop complexity penalize statically compiled kernels the most. In single-precision, JoltQC slightly outperforms TeraChem on the small basis set and runs twice as fast on def2-TZVPP. 
 Both TeraChem and JoltQC still employ double precision for eigenvalue decomposition and DIIS. On the A10-24G, which has only 1/32 the FP64 throughput of its FP32 units, these FP64 linear-algebra routines consume a disproportionate fraction of the total SCF time, diminishing the effective speedup from FP32 integral evaluation. Isolating the JK kernel on the A10-24G with FP32 arithmetic, JoltQC is approximately three times faster than TeraChem across the benchmark systems (Table \ref{tab:benchmark_fp64}, JK/cycle
column).

\begin{table*}
\centering
\small
\begin{tabular}{l|l|r|r|r|r|r|r}
\multirow{2}{*}{Molecule} & \multirow{2}{*}{Protocol} &
\multicolumn{2}{c|}{GPU4PySCF}
& \multicolumn{2}{c|}{TeraChem}
& \multicolumn{2}{c}{JoltQC}\\ \cline{3-8}
& & JK/cycle & SCF & JK/cycle & SCF & JK/cycle & SCF\\
\hline
gly30 & 6-31G*/A100/FP64 & 1.9 s & 28.1 s & 1.3 s & 19.1 s & 0.8 s & 10.4 s \\
gly30 & def2-TZVPP/A100/FP64 & 78.4 s & 1,125.7 s & 91.4 s & 1,055.1 s & 18.8 s & 268.1 s \\
gly30 & 6-31G*/A10/FP32 & - & - & 1.0 s & 23.1 s & 0.7 s & 16.5 s \\
gly30 & def2-TZVPP/A10/FP32 & - & - & 63.5 s & 870.8 s & 20.0 s & 406.1 s \\
\hline
valinomycin & 6-31G*/A100/FP64 & 2.4 s & 34.5 s & 1.9 s & 24.2 s & 1.1 s & 14.8 s \\
valinomycin & def2-TZVPP/A100/FP64 & 113.0 s & 1,579.6 s & 162.0 s & 1,641.6 s & 30.0 s & 419.2 s \\
valinomycin & cc-pVTZ/A100/FP64 & 96.0 s & 1,349.5 s & 140.0 s & 1,534.5 s & 24.0 s & 330.6 s \\
\hline
tamoxifen & cc-pVQZ/A100/FP64 & 152.0 s & 1,977.4 s & - & - & 21.4 s & 278.9 s \\
raffinose & cc-pVQZ/A100/FP64 & 188.0 s & 2,252.9 s & - & - & 28.6 s & 343.6 s \\
sphingomyelin & cc-pVQZ/A100/FP64 & 227.0 s & 2,726.0 s & - & - & 29.7 s & 356.1 s \\
\end{tabular}
\caption{\label{tab:benchmark_fp64}Benchmark among GPU4PySCF v1.4, TeraChem v1.9, and JoltQC v0.1, on A100-80G using FP64 arithmetic, and on A10-24G using FP32 arithmetic. Dashes indicate that the software does not support the required angular momentum (g functions in cc-pVQZ)}
\end{table*}

\section{Conclusion}
We have demonstrated that just-in-time compilation is a powerful paradigm for accelerating electron repulsion integral evaluation on GPUs. By making angular momentum, orbital contraction pattern, and primitive counts compile-time constants, the JIT compiler can aggressively unroll loops, eliminate branches, and optimize register allocation — yielding a 2× speedup for small basis sets (6-31G*) and up to 4× for large basis sets (def2-TZVPP) over GPU4PySCF v1.4 on an A100-80G GPU. In single precision on an A10-24G, JoltQC is approximately three times faster than TeraChem v1.9 for the JK kernels.

Beyond performance, JIT compilation fundamentally changes the development model: the core NVRTC implementation spans only 1,000 lines for both algorithms, compared to $\sim$20,000 lines in GPU4PySCF v1.4. This compactness is not incidental — it is a direct consequence of delegating specialization to the compiler rather than encoding it by hand. The resulting code is easier to maintain, extend, and adapt to new hardware. We expect JIT compilation to become a standard tool in the quantum-chemistry software stack as GPU architectures continue to diversify.

We release JoltQC as an open-source library to serve as both a practical tool and a foundation for applying JIT compilation more broadly in quantum chemistry. JoltQC supports single- and double-precision arithmetic, accepts multiple density matrices, and integrates seamlessly with GPU4PySCF. 

\section{Future Work and Limitations}
\label{sec:future}
{\bf High level abstractions of JIT}. Although our current implementation relies heavily on handwritten CUDA kernels, it would be preferable to express these routines in a high-level language such as Triton \cite{triton}, Numba \cite{numba}, or JAX \cite{jax2018github}. These frameworks offer richer support for kernel fusion, making it easier to embed quantum-chemistry operations within machine-learning pipelines. For example, one could fuse the on-the-fly computation of integrals with subsequent matrix–matrix multiplications (not only contraction with density matrices), yielding more efficient end-to-end workflows. Further, one could write a customized kernel for the derivatives with respect to contraction coefficients and exponents, then port to an existing auto-differentiation framework \cite{pyscfad}.

{\bf Supporting more functionalities}. A practical quantum‐chemistry workflow requires more than the acceleration of Coulomb and exchange kernels. The exchange-correlation functional in DFT, the gradients and Hessians of Coulomb, exchange and exchange-correlation terms, post-Hartree-Fock methods, and a wide array of molecular properties are all essential. By integrating JoltQC with GPU4PySCF, users can build sophisticated simulation pipelines, and any of these routines can be further accelerated on demand using JIT compilation.

{\bf Mixed precision algorithms}. This paper focuses on the performance of the double precision and single precision separately. Thanks to the Schwartz inequality, the computational load of the integrals can be naturally split into FP64 calculations and FP32 calculations without loss of accuracy theoretically. Other research efforts \cite{terachem-mixed,quick-mixed} show that 90\% of the computations can be in FP32. 
A straightforward mixed-precision strategy follows directly from the Schwarz inequality: shell quartets whose integral upper bound exceeds the FP32 representable range are computed in FP64, while the remaining ~90\% are computed in FP32. JoltQC's JIT framework makes this trivial to implement — the precision is already a template parameter (Table I), so both FP32 and FP64 kernels are compiled and dispatched based on the screening estimate. A key subtlety is that even when the final integral value lies within FP32 representable range, the intermediate arithmetic — particularly the recursive evaluation of Rys polynomials and the accumulation of primitive contributions — may involve catastrophic cancellation or large intermediate values that require FP64. A rigorous mixed-precision strategy must therefore identify which computational steps can safely use FP32 and which must remain in FP64 to avoid error accumulation. This is a nontrivial compiler-level analysis that we leave to future work.

{\bf Screening techniques}. The intrinsic scalings of the Coulomb and exchange terms are different. The number of significant Coulomb terms increases quadratically, while the number of significant exchange terms increases linearly with respect to the system size. It is more efficient to evaluate the Coulomb and exchange terms separately. In some quantum chemistry packages such as TeraChem, the Coulomb terms are efficiently evaluated with J-engine \cite{j_engine_original}. Besides the Schwartz inequality, which gives a rigorous upper bound, other estimates can further eliminate the unnecessary calculations. For example, the MBIE \cite{MBIE}, QQR \cite{qqr}, and CSAM \cite{CSAM}. MBIE provides a rigorous upperbound, but it is expensive to calculate. 
QQR and CSAM exploit distance dependence in their estimates, offering tighter bounds for large systems at the cost of reduced rigor compared to the Schwarz inequality. Integrating these screening strategies into JoltQC is straightforward: because the screening is performed on the host before kernel dispatch, it is independent of the JIT-compiled kernels and can be swapped without recompilation.

\section*{Author Contributions}
Xiaojie Wu led the conceptualization of JIT algorithms and developed JoltQC v0.1, and wrote the initial draft.
Qiming Sun contributed the improvements of the JK kernels v1.0 to v1.4 in GPU4PySCF, which are not documented elsewhere.
Yuanheng Wang explored single-precision algorithms in GPU4PySCF, discussed JIT algorithms in JoltQC v0.1, and wrote the manuscript.

\section*{Acknowledgments}
The authors appreciate the support of Dr. Wen Yan and the ByteDance Seed management team. Special thanks are also extended to Dr. Zhichen Pu for his discussions.  The authors also thank the PySCF community for providing valuable tools that contributed to this work, especially the GPU4PySCF contributors.

\clearpage

\bibliographystyle{plainnat}
\bibliography{main}

\begin{thebibliography}{35}
\providecommand{\natexlab}[1]{#1}
\providecommand{\url}[1]{\texttt{#1}}
\expandafter\ifx\csname urlstyle\endcsname\relax
  \providecommand{\doi}[1]{doi: #1}\else
  \providecommand{\doi}{doi: \begingroup \urlstyle{rm}\Url}\fi

\bibitem[Abadi et~al.(2015)Abadi, Agarwal, Barham, Brevdo, Chen, Citro,
  Corrado, Davis, Dean, Devin, Ghemawat, Goodfellow, Harp, Irving, Isard, Jia,
  Jozefowicz, Kaiser, Kudlur, Levenberg, Man\'{e}, Monga, Moore, Murray, Olah,
  Schuster, Shlens, Steiner, Sutskever, Talwar, Tucker, Vanhoucke, Vasudevan,
  Vi\'{e}gas, Vinyals, Warden, Wattenberg, Wicke, Yu, and
  Zheng]{tensorflow2015-whitepaper}
Mart\'{i}n Abadi, Ashish Agarwal, Paul Barham, Eugene Brevdo, Zhifeng Chen,
  Craig Citro, Greg~S. Corrado, Andy Davis, Jeffrey Dean, Matthieu Devin,
  Sanjay Ghemawat, Ian Goodfellow, Andrew Harp, Geoffrey Irving, Michael Isard,
  Yangqing Jia, Rafal Jozefowicz, Lukasz Kaiser, Manjunath Kudlur, Josh
  Levenberg, Dandelion Man\'{e}, Rajat Monga, Sherry Moore, Derek Murray, Chris
  Olah, Mike Schuster, Jonathon Shlens, Benoit Steiner, Ilya Sutskever, Kunal
  Talwar, Paul Tucker, Vincent Vanhoucke, Vijay Vasudevan, Fernanda Vi\'{e}gas,
  Oriol Vinyals, Pete Warden, Martin Wattenberg, Martin Wicke, Yuan Yu, and
  Xiaoqiang Zheng.
\newblock {TensorFlow}: Large-scale machine learning on heterogeneous systems,
  2015.
\newblock URL \url{https://www.tensorflow.org/}.
\newblock Software available from tensorflow.org.

\bibitem[Asadchev and Valeev(2024)]{LibintX}
Andrey Asadchev and Edward~F. Valeev.
\newblock 3-center and 4-center 2-particle gaussian ao integrals on modern
  accelerated processors.
\newblock \emph{The Journal of Chemical Physics}, 160\penalty0 (24):\penalty0
  244109, 06 2024.
\newblock ISSN 0021-9606.
\newblock \doi{10.1063/5.0217001}.
\newblock URL \url{https://doi.org/10.1063/5.0217001}.

\bibitem[Asadchev et~al.(2010)Asadchev, Allada, Felder, Bode, Gordon, and
  Windus]{1qnt_first}
Andrey Asadchev, Veerendra Allada, Jacob Felder, Brett~M. Bode, Mark~S. Gordon,
  and Theresa~L. Windus.
\newblock Uncontracted rys quadrature implementation of up to g functions on
  graphical processing units.
\newblock \emph{Journal of Chemical Theory and Computation}, 6\penalty0
  (3):\penalty0 696--704, 2010.
\newblock \doi{10.1021/ct9005079}.
\newblock URL \url{https://doi.org/10.1021/ct9005079}.
\newblock PMID: 26613300.

\bibitem[Becke(2014)]{DFT_overview}
Axel~D. Becke.
\newblock Perspective: Fifty years of density-functional theory in chemical
  physics.
\newblock \emph{The Journal of Chemical Physics}, 140\penalty0 (18):\penalty0
  18A301, 04 2014.
\newblock ISSN 0021-9606.
\newblock \doi{10.1063/1.4869598}.
\newblock URL \url{https://doi.org/10.1063/1.4869598}.

\bibitem[Bradbury et~al.(2018)Bradbury, Frostig, Hawkins, Johnson, Leary,
  Maclaurin, Necula, Paszke, Vander{P}las, Wanderman-{M}ilne, and
  Zhang]{jax2018github}
James Bradbury, Roy Frostig, Peter Hawkins, Matthew~James Johnson, Chris Leary,
  Dougal Maclaurin, George Necula, Adam Paszke, Jake Vander{P}las, Skye
  Wanderman-{M}ilne, and Qiao Zhang.
\newblock {JAX}: composable transformations of {P}ython+{N}um{P}y programs,
  2018.
\newblock URL \url{http://github.com/jax-ml/jax}.

\bibitem[Dill and Pople(1975)]{6-31g}
James~D. Dill and John~A. Pople.
\newblock Self-consistent molecular orbital methods. xv. extended gaussian-type
  basis sets for lithium, beryllium, and boron.
\newblock \emph{J. Chem. Phys.}, 62:\penalty0 2921--2923, 1975.
\newblock \doi{10.1063/1.430801}.

\bibitem[Dupuis et~al.(1976)Dupuis, Rys, and King]{rys_first}
Michel Dupuis, John Rys, and Harry~F. King.
\newblock Evaluation of molecular integrals over gaussian basis functions.
\newblock \emph{The Journal of Chemical Physics}, 65\penalty0 (1):\penalty0
  111--116, 07 1976.
\newblock ISSN 0021-9606.
\newblock \doi{10.1063/1.432807}.
\newblock URL \url{https://doi.org/10.1063/1.432807}.

\bibitem[Epifanovsky et~al.(2021)Epifanovsky, Gilbert, Feng, Lee, Mao,
  Mardirossian, Pokhilko, White, Coons, Dempwolff, Gan, Hait, Horn, Jacobson,
  Kaliman, Kussmann, Lange, Lao, Levine, Liu, McKenzie, Morrison, Nanda,
  Plasser, Rehn, Vidal, You, Zhu, Alam, Albrecht, Aldossary, Alguire, Andersen,
  Athavale, Barton, Begam, Behn, Bellonzi, Bernard, Berquist, Burton, Carreras,
  Carter-Fenk, Chakraborty, Chien, Closser, Cofer-Shabica, Dasgupta,
  de~Wergifosse, Deng, Diedenhofen, Do, Ehlert, Fang, Fatehi, Feng, Friedhoff,
  Gayvert, Ge, Gidofalvi, Goldey, Gomes, González-Espinoza, Gulania, Gunina,
  Hanson-Heine, Harbach, Hauser, Herbst, Hernández~Vera, Hodecker, Holden,
  Houck, Huang, Hui, Huynh, Ivanov, Jász, Ji, Jiang, Kaduk, Kähler,
  Khistyaev, Kim, Kis, Klunzinger, Koczor-Benda, Koh, Kosenkov, Koulias,
  Kowalczyk, Krauter, Kue, Kunitsa, Kus, Ladjánszki, Landau, Lawler,
  Lefrancois, Lehtola, Li, Li, Liang, Liebenthal, Lin, Lin, Liu, Liu,
  Loipersberger, Luenser, Manjanath, Manohar, Mansoor, Manzer, Mao, Marenich,
  Markovich, Mason, Maurer, McLaughlin, Menger, Mewes, Mewes, Morgante,
  Mullinax, Oosterbaan, Paran, Paul, Paul, Pavošević, Pei, Prager, Proynov,
  Rák, Ramos-Cordoba, Rana, Rask, Rettig, Richard, Rob, Rossomme, Scheele,
  Scheurer, Schneider, Sergueev, Sharada, Skomorowski, Small, Stein, Su,
  Sundstrom, Tao, Thirman, Tornai, Tsuchimochi, Tubman, Veccham, Vydrov,
  Wenzel, Witte, Yamada, Yao, Yeganeh, Yost, Zech, Zhang, Zhang, Zhang, Zuev,
  Aspuru-Guzik, Bell, Besley, Bravaya, Brooks, Casanova, Chai, Coriani, Cramer,
  Cserey, DePrince, DiStasio, Dreuw, Dunietz, Furlani, Goddard,
  Hammes-Schiffer, Head-Gordon, Hehre, Hsu, Jagau, Jung, Klamt, Kong,
  Lambrecht, Liang, Mayhall, McCurdy, Neaton, Ochsenfeld, Parkhill, Peverati,
  Rassolov, Shao, Slipchenko, Stauch, Steele, Subotnik, Thom, Tkatchenko,
  Truhlar, Van~Voorhis, Wesolowski, Whaley, Woodcock, Zimmerman, Faraji, Gill,
  Head-Gordon, Herbert, and Krylov]{qchem5}
Evgeny Epifanovsky, Andrew T.~B. Gilbert, Xintian Feng, Joonho Lee, Yuezhi Mao,
  Narbe Mardirossian, Pavel Pokhilko, Alec~F. White, Marc~P. Coons, Adrian~L.
  Dempwolff, Zhengting Gan, Diptarka Hait, Paul~R. Horn, Leif~D. Jacobson, Ilya
  Kaliman, Jörg Kussmann, Adrian~W. Lange, Ka~Un Lao, Daniel~S. Levine, Jie
  Liu, Simon~C. McKenzie, Adrian~F. Morrison, Kaushik~D. Nanda, Felix Plasser,
  Dirk~R. Rehn, Marta~L. Vidal, Zhi-Qiang You, Ying Zhu, Bushra Alam,
  Benjamin~J. Albrecht, Abdulrahman Aldossary, Ethan Alguire, Josefine~H.
  Andersen, Vishikh Athavale, Dennis Barton, Khadiza Begam, Andrew Behn, Nicole
  Bellonzi, Yves~A. Bernard, Eric~J. Berquist, Hugh G.~A. Burton, Abel
  Carreras, Kevin Carter-Fenk, Romit Chakraborty, Alan~D. Chien, Kristina~D.
  Closser, Vale Cofer-Shabica, Saswata Dasgupta, Marc de~Wergifosse, Jia Deng,
  Michael Diedenhofen, Hainam Do, Sebastian Ehlert, Po-Tung Fang, Shervin
  Fatehi, Qingguo Feng, Triet Friedhoff, James Gayvert, Qinghui Ge, Gergely
  Gidofalvi, Matthew Goldey, Joe Gomes, Cristina~E. González-Espinoza, Sahil
  Gulania, Anastasia~O. Gunina, Magnus W.~D. Hanson-Heine, Phillip H.~P.
  Harbach, Andreas Hauser, Michael~F. Herbst, Mario Hernández~Vera, Manuel
  Hodecker, Zachary~C. Holden, Shannon Houck, Xunkun Huang, Kerwin Hui, Bang~C.
  Huynh, Maxim Ivanov, Ádám Jász, Hyunjun Ji, Hanjie Jiang, Benjamin Kaduk,
  Sven Kähler, Kirill Khistyaev, Jaehoon Kim, Gergely Kis, Phil Klunzinger,
  Zsuzsanna Koczor-Benda, Joong~Hoon Koh, Dimitri Kosenkov, Laura Koulias, Tim
  Kowalczyk, Caroline~M. Krauter, Karl Kue, Alexander Kunitsa, Thomas Kus,
  István Ladjánszki, Arie Landau, Keith~V. Lawler, Daniel Lefrancois, Susi
  Lehtola, Run~R. Li, Yi-Pei Li, Jiashu Liang, Marcus Liebenthal, Hung-Hsuan
  Lin, You-Sheng Lin, Fenglai Liu, Kuan-Yu Liu, Matthias Loipersberger, Arne
  Luenser, Aaditya Manjanath, Prashant Manohar, Erum Mansoor, Sam~F. Manzer,
  Shan-Ping Mao, Aleksandr~V. Marenich, Thomas Markovich, Stephen Mason,
  Simon~A. Maurer, Peter~F. McLaughlin, Maximilian F. S.~J. Menger, Jan-Michael
  Mewes, Stefanie~A. Mewes, Pierpaolo Morgante, J.~Wayne Mullinax, Katherine~J.
  Oosterbaan, Garrette Paran, Alexander~C. Paul, Suranjan~K. Paul, Fabijan
  Pavošević, Zheng Pei, Stefan Prager, Emil~I. Proynov, Ádám Rák, Eloy
  Ramos-Cordoba, Bhaskar Rana, Alan~E. Rask, Adam Rettig, Ryan~M. Richard,
  Fazle Rob, Elliot Rossomme, Tarek Scheele, Maximilian Scheurer, Matthias
  Schneider, Nickolai Sergueev, Shaama~M. Sharada, Wojciech Skomorowski,
  David~W. Small, Christopher~J. Stein, Yu-Chuan Su, Eric~J. Sundstrom, Zhen
  Tao, Jonathan Thirman, Gábor~J. Tornai, Takashi Tsuchimochi, Norm~M. Tubman,
  Srimukh~Prasad Veccham, Oleg Vydrov, Jan Wenzel, Jon Witte, Atsushi Yamada,
  Kun Yao, Sina Yeganeh, Shane~R. Yost, Alexander Zech, Igor~Ying Zhang, Xing
  Zhang, Yu~Zhang, Dmitry Zuev, Alán Aspuru-Guzik, Alexis~T. Bell, Nicholas~A.
  Besley, Ksenia~B. Bravaya, Bernard~R. Brooks, David Casanova, Jeng-Da Chai,
  Sonia Coriani, Christopher~J. Cramer, György Cserey, III DePrince,
  A.~Eugene, Jr. DiStasio, Robert~A., Andreas Dreuw, Barry~D. Dunietz,
  Thomas~R. Furlani, III Goddard, William~A., Sharon Hammes-Schiffer, Teresa
  Head-Gordon, Warren~J. Hehre, Chao-Ping Hsu, Thomas-C. Jagau, Yousung Jung,
  Andreas Klamt, Jing Kong, Daniel~S. Lambrecht, WanZhen Liang, Nicholas~J.
  Mayhall, C.~William McCurdy, Jeffrey~B. Neaton, Christian Ochsenfeld, John~A.
  Parkhill, Roberto Peverati, Vitaly~A. Rassolov, Yihan Shao, Lyudmila~V.
  Slipchenko, Tim Stauch, Ryan~P. Steele, Joseph~E. Subotnik, Alex J.~W. Thom,
  Alexandre Tkatchenko, Donald~G. Truhlar, Troy Van~Voorhis, Tomasz~A.
  Wesolowski, K.~Birgitta Whaley, III Woodcock, H.~Lee, Paul~M. Zimmerman,
  Shirin Faraji, Peter M.~W. Gill, Martin Head-Gordon, John~M. Herbert, and
  Anna~I. Krylov.
\newblock Software for the frontiers of quantum chemistry: An overview of
  developments in the q-chem 5 package.
\newblock \emph{The Journal of Chemical Physics}, 155\penalty0 (8):\penalty0
  084801, 08 2021.
\newblock ISSN 0021-9606.
\newblock \doi{10.1063/5.0055522}.
\newblock URL \url{https://doi.org/10.1063/5.0055522}.

\bibitem[Flocke and Lotrich(2008)]{rys_quadrature}
N.~Flocke and V.~Lotrich.
\newblock Efficient electronic integrals and their generalized derivatives for
  object oriented implementations of electronic structure calculations.
\newblock \emph{Journal of Computational Chemistry}, 29\penalty0 (16):\penalty0
  2722--2736, 2008.
\newblock \doi{https://doi.org/10.1002/jcc.21018}.
\newblock URL \url{https://onlinelibrary.wiley.com/doi/abs/10.1002/jcc.21018}.

\bibitem[Frisch et~al.(2016)Frisch, Trucks, Schlegel, Scuseria, Robb,
  Cheeseman, Scalmani, Barone, Petersson, Nakatsuji, Li, Caricato, Marenich,
  Bloino, Janesko, Gomperts, Mennucci, Hratchian, Ortiz, Izmaylov, Sonnenberg,
  Williams-Young, Ding, Lipparini, Egidi, Goings, Peng, Petrone, Henderson,
  Ranasinghe, Zakrzewski, Gao, Rega, Zheng, Liang, Hada, Ehara, Toyota, Fukuda,
  Hasegawa, Ishida, Nakajima, Honda, Kitao, Nakai, Vreven, Throssell,
  Montgomery, Peralta, Ogliaro, Bearpark, Heyd, Brothers, Kudin, Staroverov,
  Keith, Kobayashi, Normand, Raghavachari, Rendell, Burant, Iyengar, Tomasi,
  Cossi, Millam, Klene, Adamo, Cammi, Ochterski, Martin, Morokuma, Farkas,
  Foresman, and Fox]{g16}
M.~J. Frisch, G.~W. Trucks, H.~B. Schlegel, G.~E. Scuseria, M.~A. Robb, J.~R.
  Cheeseman, G.~Scalmani, V.~Barone, G.~A. Petersson, H.~Nakatsuji, X.~Li,
  M.~Caricato, A.~V. Marenich, J.~Bloino, B.~G. Janesko, R.~Gomperts,
  B.~Mennucci, H.~P. Hratchian, J.~V. Ortiz, A.~F. Izmaylov, J.~L. Sonnenberg,
  D.~Williams-Young, F.~Ding, F.~Lipparini, F.~Egidi, J.~Goings, B.~Peng,
  A.~Petrone, T.~Henderson, D.~Ranasinghe, V.~G. Zakrzewski, J.~Gao, N.~Rega,
  G.~Zheng, W.~Liang, M.~Hada, M.~Ehara, K.~Toyota, R.~Fukuda, J.~Hasegawa,
  M.~Ishida, T.~Nakajima, Y.~Honda, O.~Kitao, H.~Nakai, T.~Vreven,
  K.~Throssell, J.~A. Montgomery, {Jr.}, J.~E. Peralta, F.~Ogliaro, M.~J.
  Bearpark, J.~J. Heyd, E.~N. Brothers, K.~N. Kudin, V.~N. Staroverov, T.~A.
  Keith, R.~Kobayashi, J.~Normand, K.~Raghavachari, A.~P. Rendell, J.~C.
  Burant, S.~S. Iyengar, J.~Tomasi, M.~Cossi, J.~M. Millam, M.~Klene, C.~Adamo,
  R.~Cammi, J.~W. Ochterski, R.~L. Martin, K.~Morokuma, O.~Farkas, J.~B.
  Foresman, and D.~J. Fox.
\newblock Gaussian˜16 {R}evision {C}.01, 2016.
\newblock Gaussian Inc. Wallingford CT.

\bibitem[Head‐Gordon and Pople(1988)]{HGP_original}
Martin Head‐Gordon and John~A. Pople.
\newblock A method for two‐electron gaussian integral and integral derivative
  evaluation using recurrence relations.
\newblock \emph{The Journal of Chemical Physics}, 89\penalty0 (9):\penalty0
  5777--5786, 11 1988.
\newblock ISSN 0021-9606.
\newblock \doi{10.1063/1.455553}.
\newblock URL \url{https://doi.org/10.1063/1.455553}.

\bibitem[Helgaker et~al.(2000)Helgaker, Jørgensen, and
  Olsen]{purple_book_chapter_basis}
Trygve Helgaker, Poul Jørgensen, and Jeppe Olsen.
\newblock \emph{Atomic Basis Functions}, chapter~6, pages 201--255.
\newblock John Wiley \& Sons, Ltd, 2000.
\newblock ISBN 9781119019572.
\newblock \doi{https://doi.org/10.1002/9781119019572.ch6}.
\newblock URL
  \url{https://onlinelibrary.wiley.com/doi/abs/10.1002/9781119019572.ch6}.

\bibitem[Helgaker et~al.(2013)Helgaker, Jorgensen, and Olsen]{purple_book}
Trygve Helgaker, Poul Jorgensen, and Jeppe Olsen.
\newblock \emph{Molecular electronic-structure theory}.
\newblock John Wiley \& Sons, 2013.

\bibitem[Lam et~al.(2015)Lam, Pitrou, and Seibert]{numba}
Siu~Kwan Lam, Antoine Pitrou, and Stanley Seibert.
\newblock Numba: a llvm-based python jit compiler.
\newblock In \emph{Proceedings of the Second Workshop on the LLVM Compiler
  Infrastructure in HPC}, LLVM '15, New York, NY, USA, 2015. Association for
  Computing Machinery.
\newblock ISBN 9781450340052.
\newblock \doi{10.1145/2833157.2833162}.
\newblock URL \url{https://doi.org/10.1145/2833157.2833162}.

\bibitem[Lambrecht et~al.(2005)Lambrecht, Doser, and Ochsenfeld]{MBIE}
Daniel~S. Lambrecht, Bernd Doser, and Christian Ochsenfeld.
\newblock Rigorous integral screening for electron correlation methods.
\newblock \emph{The Journal of Chemical Physics}, 123\penalty0 (18):\penalty0
  184102, 11 2005.
\newblock ISSN 0021-9606.
\newblock \doi{10.1063/1.2079987}.
\newblock URL \url{https://doi.org/10.1063/1.2079987}.

\bibitem[Lattner and Adve(2004)]{llvm}
C.~Lattner and V.~Adve.
\newblock Llvm: a compilation framework for lifelong program analysis \&
  transformation.
\newblock In \emph{International Symposium on Code Generation and Optimization,
  2004. CGO 2004.}, pages 75--86, 2004.
\newblock \doi{10.1109/CGO.2004.1281665}.

\bibitem[Li et~al.(2025)Li, Sun, Zhang, and Chan]{GPU4PySCF1}
Rui Li, Qiming Sun, Xing Zhang, and Garnet Kin-Lic Chan.
\newblock Introducing gpu acceleration into the python-based simulations of
  chemistry framework.
\newblock \emph{The Journal of Physical Chemistry A}, 129\penalty0
  (5):\penalty0 1459--1468, 2025.
\newblock \doi{10.1021/acs.jpca.4c05876}.
\newblock URL \url{https://doi.org/10.1021/acs.jpca.4c05876}.
\newblock PMID: 39846468.

\bibitem[Luehr et~al.(2011)Luehr, Ufimtsev, and Martínez]{terachem-mixed}
Nathan Luehr, Ivan~S. Ufimtsev, and Todd~J. Martínez.
\newblock Dynamic precision for electron repulsion integral evaluation on
  graphical processing units (gpus).
\newblock \emph{Journal of Chemical Theory and Computation}, 7\penalty0
  (4):\penalty0 949--954, 2011.
\newblock \doi{10.1021/ct100701w}.
\newblock URL \url{https://doi.org/10.1021/ct100701w}.
\newblock PMID: 26606344.

\bibitem[Maurer et~al.(2012)Maurer, Lambrecht, Flaig, and Ochsenfeld]{qqr}
Simon~A. Maurer, Daniel~S. Lambrecht, Denis Flaig, and Christian Ochsenfeld.
\newblock Distance-dependent schwarz-based integral estimates for two-electron
  integrals: Reliable tightness vs. rigorous upper bounds.
\newblock \emph{The Journal of Chemical Physics}, 136\penalty0 (14):\penalty0
  144107, 04 2012.
\newblock ISSN 0021-9606.
\newblock \doi{10.1063/1.3693908}.
\newblock URL \url{https://doi.org/10.1063/1.3693908}.

\bibitem[McMurchie and Davidson(1978)]{MD_original}
Larry~E McMurchie and Ernest~R Davidson.
\newblock One- and two-electron integrals over cartesian gaussian functions.
\newblock \emph{Journal of Computational Physics}, 26\penalty0 (2):\penalty0
  218--231, 1978.
\newblock ISSN 0021-9991.
\newblock \doi{https://doi.org/10.1016/0021-9991(78)90092-X}.
\newblock URL
  \url{https://www.sciencedirect.com/science/article/pii/002199917890092X}.

\bibitem[Miao and Merz(2013)]{quick-mixed}
Yipu Miao and Kenneth M.~Jr. Merz.
\newblock Acceleration of electron repulsion integral evaluation on graphics
  processing units via use of recurrence relations.
\newblock \emph{Journal of Chemical Theory and Computation}, 9\penalty0
  (2):\penalty0 965--976, 2013.
\newblock \doi{10.1021/ct300754n}.
\newblock URL \url{https://doi.org/10.1021/ct300754n}.
\newblock PMID: 26588740.

\bibitem[Neese(2022)]{ORCA5}
F.~Neese.
\newblock Software update: the orca program system, version 5.0.
\newblock \emph{WIRES Comput. Molec. Sci.}, 12\penalty0 (1):\penalty0 e1606,
  2022.
\newblock \doi{10.1002/wcms.1606}.

\bibitem[{NVIDIA Corporation}(2025)]{nvrtc}
{NVIDIA Corporation}.
\newblock \emph{{NVRTC}: NVIDIA Runtime Compilation Library}.
\newblock NVIDIA Corporation, 2025.
\newblock URL \url{https://docs.nvidia.com/cuda/nvrtc/index.html}.
\newblock Accessed: 2025-06-17.

\bibitem[Okuta et~al.(2017)Okuta, Unno, Nishino, Hido, and
  Loomis]{cupy_learningsys2017}
Ryosuke Okuta, Yuya Unno, Daisuke Nishino, Shohei Hido, and Crissman Loomis.
\newblock Cupy: A numpy-compatible library for nvidia gpu calculations.
\newblock In \emph{Proceedings of Workshop on Machine Learning Systems
  (LearningSys) in The Thirty-first Annual Conference on Neural Information
  Processing Systems (NIPS)}, 2017.
\newblock URL \url{http://learningsys.org/nips17/assets/papers/paper_16.pdf}.

\bibitem[Paszke et~al.(2019)Paszke, Gross, Massa, Lerer, Bradbury, Chanan,
  Killeen, Lin, Gimelshein, Antiga, Desmaison, K\"{o}pf, Yang, DeVito, Raison,
  Tejani, Chilamkurthy, Steiner, Fang, Bai, and Chintala]{pytorch}
Adam Paszke, Sam Gross, Francisco Massa, Adam Lerer, James Bradbury, Gregory
  Chanan, Trevor Killeen, Zeming Lin, Natalia Gimelshein, Luca Antiga, Alban
  Desmaison, Andreas K\"{o}pf, Edward Yang, Zach DeVito, Martin Raison, Alykhan
  Tejani, Sasank Chilamkurthy, Benoit Steiner, Lu~Fang, Junjie Bai, and Soumith
  Chintala.
\newblock \emph{PyTorch: an imperative style, high-performance deep learning
  library}.
\newblock Curran Associates Inc., Red Hook, NY, USA, 2019.

\bibitem[Pritchard et~al.(2019)Pritchard, Altarawy, Didier, Gibsom, and
  Windus]{basissetexchange}
Benjamin~P. Pritchard, Doaa Altarawy, Brett Didier, Tara~D. Gibsom, and
  Theresa~L. Windus.
\newblock A new basis set exchange: An open, up-to-date resource for the
  molecular sciences community.
\newblock \emph{J. Chem. Inf. Model.}, 59:\penalty0 4814--4820, 2019.
\newblock \doi{10.1021/acs.jcim.9b00725}.

\bibitem[Sun(2015)]{libcint}
Qiming Sun.
\newblock Libcint: An efficient general integral library for gaussian basis
  functions.
\newblock \emph{Journal of Computational Chemistry}, 36\penalty0 (22):\penalty0
  1664--1671, 2015.
\newblock \doi{https://doi.org/10.1002/jcc.23981}.
\newblock URL \url{https://onlinelibrary.wiley.com/doi/abs/10.1002/jcc.23981}.

\bibitem[Thompson and Ochsenfeld(2017)]{CSAM}
Travis~H. Thompson and Christian Ochsenfeld.
\newblock Distance-including rigorous upper bounds and tight estimates for
  two-electron integrals over long- and short-range operators.
\newblock \emph{The Journal of Chemical Physics}, 147\penalty0 (14):\penalty0
  144101, 10 2017.
\newblock ISSN 0021-9606.
\newblock \doi{10.1063/1.4994190}.
\newblock URL \url{https://doi.org/10.1063/1.4994190}.

\bibitem[Tillet et~al.(2019)Tillet, Kung, and Cox]{triton}
Philippe Tillet, H.~T. Kung, and David Cox.
\newblock Triton: an intermediate language and compiler for tiled neural
  network computations.
\newblock In \emph{Proceedings of the 3rd ACM SIGPLAN International Workshop on
  Machine Learning and Programming Languages}, MAPL 2019, page 10–19, New
  York, NY, USA, 2019. Association for Computing Machinery.
\newblock ISBN 9781450367196.
\newblock \doi{10.1145/3315508.3329973}.
\newblock URL \url{https://doi.org/10.1145/3315508.3329973}.

\bibitem[Vallejo et~al.(2023)Vallejo, Barca, and and]{Gamess_f}
Jorge Luis~Galvez Vallejo, Giuseppe~M.J. Barca, and Mark S.~Gordon and.
\newblock High-performance gpu-accelerated evaluation of electron repulsion
  integrals.
\newblock \emph{Molecular Physics}, 121\penalty0 (9-10):\penalty0 e2112987,
  2023.
\newblock \doi{10.1080/00268976.2022.2112987}.
\newblock URL \url{https://doi.org/10.1080/00268976.2022.2112987}.

\bibitem[Wang et~al.(2024)Wang, Hait, Johnson, Fajen, Zhang, Guerrero, and
  Martínez]{terachem2024}
Yuanheng Wang, Diptarka Hait, K.~Grace Johnson, O.~Jonathan Fajen,
  Juncheng~Harry Zhang, Rubén~D. Guerrero, and Todd~J. Martínez.
\newblock Extending gpu-accelerated gaussian integrals in the terachem software
  package to f type orbitals: Implementation and applications.
\newblock \emph{The Journal of Chemical Physics}, 161\penalty0 (17):\penalty0
  174118, 11 2024.
\newblock ISSN 0021-9606.
\newblock \doi{10.1063/5.0233523}.
\newblock URL \url{https://doi.org/10.1063/5.0233523}.

\bibitem[Weigend and Ahlrichs(2005)]{def2-tzvx}
Florian Weigend and Reinhart Ahlrichs.
\newblock Balanced basis sets of split valence, triple zeta valence and
  quadruple zeta valence quality for h to rn: Design and assessment of
  accuracy.
\newblock \emph{Phys. Chem. Chem. Phys.}, 7:\penalty0 3297, 2005.
\newblock \doi{10.1039/b508541a}.

\bibitem[White and Head‐Gordon(1996)]{j_engine_original}
Christopher~A. White and Martin Head‐Gordon.
\newblock A j matrix engine for density functional theory calculations.
\newblock \emph{The Journal of Chemical Physics}, 104\penalty0 (7):\penalty0
  2620--2629, 02 1996.
\newblock ISSN 0021-9606.
\newblock \doi{10.1063/1.470986}.
\newblock URL \url{https://doi.org/10.1063/1.470986}.

\bibitem[Wu et~al.(2025)Wu, Sun, Pu, Zheng, Ma, Yan, Xia, Wu, Huo, Li, Ren,
  Gong, Zhang, and Gao]{GPU4PySCF2}
Xiaojie Wu, Qiming Sun, Zhichen Pu, Tianze Zheng, Wenzhi Ma, Wen Yan, Yu~Xia,
  Zhengxiao Wu, Mian Huo, Xiang Li, Weiluo Ren, Sheng Gong, Yumin Zhang, and
  Weihao Gao.
\newblock Enhancing gpu-acceleration in the python-based simulations of
  chemistry frameworks.
\newblock \emph{WIREs Computational Molecular Science}, 15\penalty0
  (2):\penalty0 e70008, 2025.
\newblock \doi{https://doi.org/10.1002/wcms.70008}.
\newblock URL
  \url{https://wires.onlinelibrary.wiley.com/doi/abs/10.1002/wcms.70008}.
\newblock e70008 CMS-1146.R2.

\bibitem[Zhang and Chan(2022)]{pyscfad}
Xing Zhang and Garnet Kin-Lic Chan.
\newblock Differentiable quantum chemistry with pyscf for molecules and
  materials at the mean-field level and beyond.
\newblock \emph{The Journal of Chemical Physics}, 157\penalty0 (20):\penalty0
  204801, 11 2022.
\newblock ISSN 0021-9606.
\newblock \doi{10.1063/5.0118200}.
\newblock URL \url{https://doi.org/10.1063/5.0118200}.

\end{thebibliography}

\clearpage

\beginappendix

\section{TeraChem input sample}
In this paper, we use the following sample TeraChem script. The coordinates, basis, and precision arguments are changed correspondingly.
\label{sec:terachem_input}
\begin{verbatim}
# A sample of TeraChem script
run energy
method hf
basis 6-31gs
precision single
coordinates gly30.xyz
charge 0
timings yes
threcl 1e-13
threex 1e-13
guess sad
xtol 1e-6
purify no
fock incremental
\end{verbatim}

\section{Analyze the bottlenecks of 1q1t algorithm and 1qnt algorithm}
\label{sec:appendix_analyze}
We benchmark the 1q1t algorithm on the gly120 system with the following customized basis set in NWChem format \cite{basissetexchange},
\begin{verbatim}
X    S
      0.2700058226E+00      1
      0.2700058226E+00      1
      0.2700058226E+00      1
      0.2700058226E+00      1
      0.2700058226E+00      1
      0.2700058226E+00      1
\end{verbatim}
The same basis set is assigned to all types of elements, for profiling purpose only. The number of primitives varies in the tests. The 1q1t algorithm is largely memory bound when the number of primitives is small. For a (ss|ss) kernel with uncontracted orbitals, when switching from FP64 to FP32, 
 the FP32-over-FP64 speedup is only 2×.
The computing intensity increases rapidly with the number of primitives. When the numbers of primitives of all the shells are 4, the FP32 to FP64 speedup achieves the theoretical peak performance of A10-24G, which has 32:1 FP64 / FP32 compute units. However, when the number of primitives is greater than 5, we observed register spill onto local memory. This is essentially due to the compiler (NVRTC/NVCC) not unrolling the nested loops of primitives. There are multiple ways to manually unroll the loops, but for simplicity of the code, we leave it for future work. A similar issue also occurs in the case of other patterns, such as $(fs|pp)$, $(fd|ss)$, and so on.
\begin{figure}
    \centering
    \includegraphics[width=0.7\linewidth]{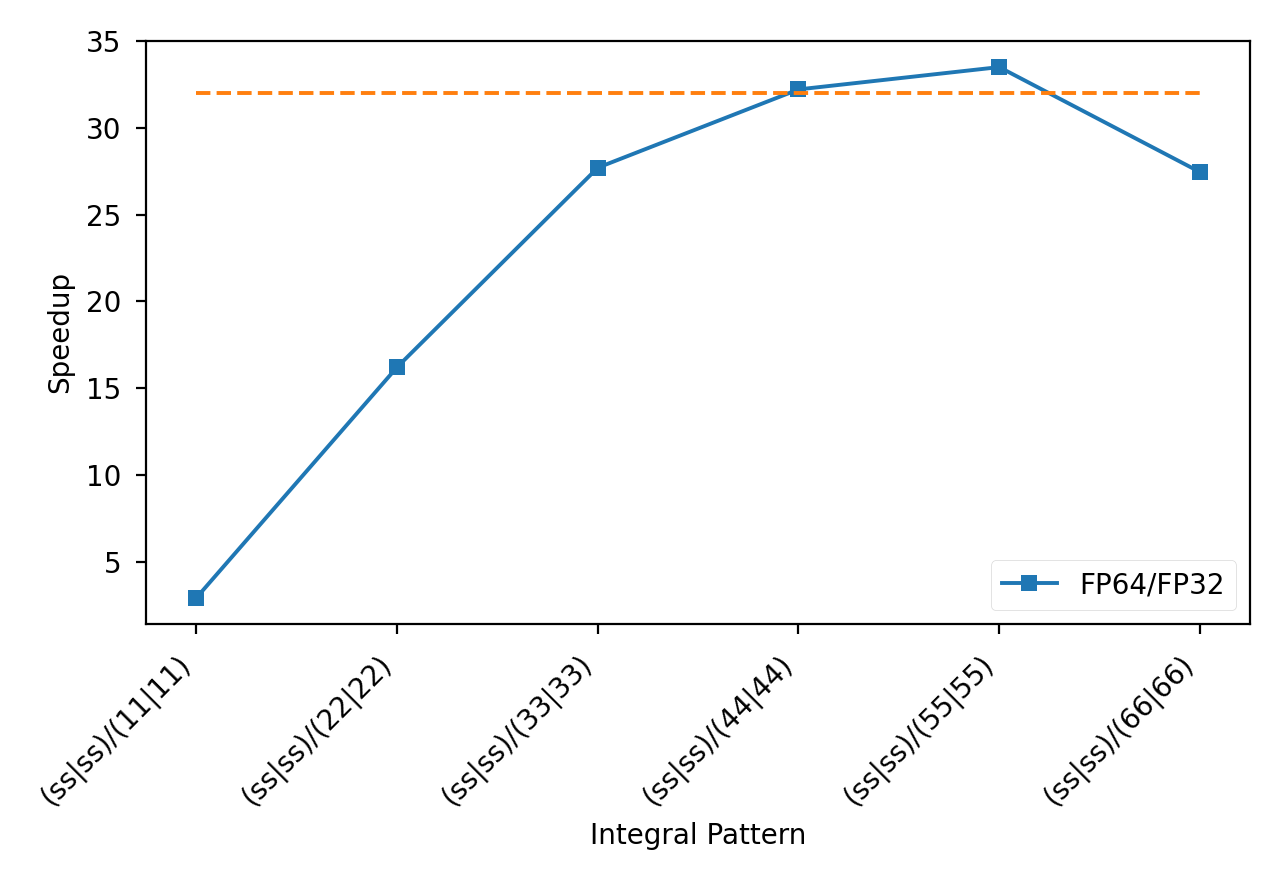}
    \caption{Speedup of FP32/FP64 JK calculation on A10-24G for $(ss|ss)$ with different number of primitives. The computation is performed on a gly120 system, with a customized basis set composed of s orbitals with the corresponding number of primitive only (described in main text). The orange dashed line labels the theoretical FP32/FP64 performance ratio of A10-24G, which is 32:1.}
    \label{fig:1q1t_analyze}
\end{figure}

\begin{figure}
    \centering
    \includegraphics[width=0.7\linewidth]{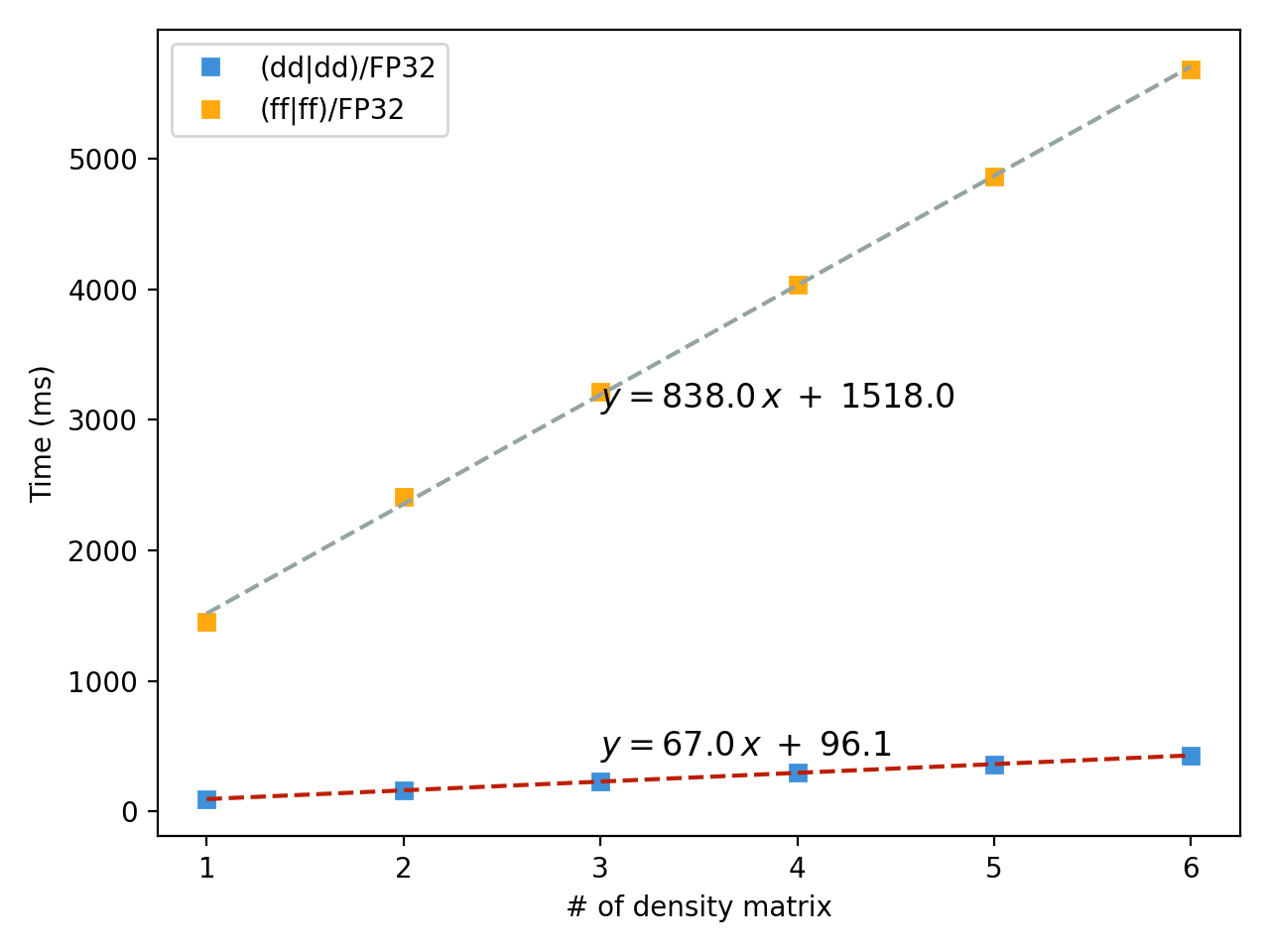}
    \caption{Timing of JK calculation of the $(dd|dd)/(11|11)$ and $(ff|ff)/(11|11)$ kernels with multiple density matrices on A10-24G. The computation is performed on a gly30 system, with a customized basis set described in the main text.}
    \label{fig:1qnt_analyze}
\end{figure}

The performance bottleneck in the 1qnt algorithm arises from two distinct phases: integral generation and contraction. In the {\bf integral generation} phase, the cost of calculating the integral values is independent of the number of density matrices. In the {\bf contraction} phase, the integral value is contracted with every density matrix, so its cost grows linearly with the matrix count. To quantify this behavior, we vary the number of density matrices and recorded the total wall-clock time (Figure \ref{fig:1qnt_analyze}). We constructed the following basis set in NWChem format for the test, which assigns the same d and f orbitals to all types of elements:
\begin{verbatim}
X    D
      1.2700058226E+00      1
X    F
      1.2700058226E+00      1
\end{verbatim}

The results show an almost perfectly linear relation between the overall cost and the number of density matrices. For the $(dd|dd)$ pattern, integral generation requires 96 ms and each contraction 67 ms; in a higher-angular-momentum case, these times rise to 1,518 ms and 837 ms, respectively. Consequently, when only one density matrix is used (e.g., in restricted HF/DFT methods), the bottleneck lies in integral generation, whereas with two or more matrices (for example, unrestricted HF/DFT methods), density matrix contraction dominates. In the latter scenario, more sophisticated contraction strategies are needed, such as tensor core-accelerated kernels.

We emphasize that this analysis is approximate: increased register pressure from handling multiple density matrices can further affect kernel performance, and those effects are not reflected in our current timing estimates.

\end{document}